\documentclass{aa}  

\usepackage{graphicx}
\usepackage{txfonts}
\usepackage{siunitx}
\usepackage{lipsum}
\usepackage{bm}
\usepackage{subcaption}         
\usepackage{lscape}             
\usepackage{placeins}           
\usepackage{color}
\usepackage{cancel}
\usepackage{natbib}
\usepackage{seqsplit}

\usepackage{hyperref}
\hypersetup{
  colorlinks=true,
  citecolor=blue,
  linkcolor=blue,
  urlcolor=blue,
}

\newcommand{\vx}{\bm{x}}

\newcommand{\Mpch}{h^{-1}\mathrm{Mpc}}
\newcommand{\volunits}{h^{-3}\mathrm{Mpc}^{3}}
\newcommand{\nunits}{h^{3}\mathrm{Mpc}^{-3}}

\newcommand{\op}{\mathcal{O}}
\newcommand{\prob}[1]{\mathcal{P}\left( #1 \right)}
\newcommand{\probdens}[1]{p\left( #1 \right)}
\newcommand{\avg}[1]{\langle #1 \rangle}
\newcommand{\abs}[1]{| #1 |}

\newcommand{\like}[1]{\mathcal{L} \left( #1 \right)}

\newcommand{\Om}{\Omega_{\rm m}}
\newcommand{\Omo}{\Omega_{\rm m, 0}}
\newcommand{\Ob}{\Omega_{\rm b}}
\newcommand{\sig}{\sigma_8}

\newcommand{\bphi}{b_{\phi}}
\newcommand{\fnl}{f_{\rm NL}^{\rm loc}}
\newcommand{\thetac}{\theta_{\rm c}}
\newcommand{\thetagf}{\theta_{\rm gf}}
\newcommand{\thetashame}{\theta_{\rm SHAMe}}

\newcommand{\mstell}{M_{\star}}
\newcommand{\mstellcut}{M_{\star}^{\rm cut}}

\newcommand{\smin}{s_{\rm min}}
\newcommand{\smax}{s_{\rm max}}
\newcommand{\xsub}{\mathcal{S}}

\newcommand{\nbar}{\bar{n}_g}
\newcommand{\vpeak}{V_{\rm peak}}
\newcommand{\vmax}{V_{\rm max}}

\newcommand{\sigmalum}{\sigma_{\mathrm{lum}}}
\newcommand{\logtmerger}{\mathrm{log}(t_{\rm merger})}

\newcommand{\fkp}{f_{k, +}}
\newcommand{\fkm}{f_{k, -}}
\newcommand{\betalum}{\beta_{\rm lum}}

\begin{document}

   \title{Small-scale priors for large-scale cosmology: the case of $b_{\phi}$}

%
%
%

   \author{M. Pinon\inst{1}\corrauth{mathilde.pinon@dipc.org}       
        \and R. E. Angulo\inst{1, 2}
        \and S. Contreras\inst{3}
        \and A. Moradinezhad Dizgah\inst{4}
        }

   \institute{Donostia International Physics Center (DIPC), Donostia-San Sebastian, Spain
   \and IKERBASQUE, Basque Foundation for Science, 48013, Bilbao, Spain
   \and Facultad de Física, Universidad de Sevilla, Campus de Reina Mercedes, Av. Reina Mercedes s/n 41012 Seville, Spain
   \and Laboratoire d'Annecy de Physique Theorique (LAPTh), CNRS/USMB, 99 Chemin de Bellevue BP110, Annecy F-74941, France}


 
  \abstract
   {Perturbative descriptions of galaxy clustering on large scales introduce free bias parameters that capture the complex physics of galaxy formation. In particular, local primordial non-Gaussianity (PNG), parametrized by $\fnl$, induces a scale-dependent contribution to the galaxy bias that is proportional to $\fnl \bphi$, where $\bphi$ quantifies how galaxy formation responds to PNG. Accurately determining $\bphi$ is therefore essential for robust constraints on $\fnl$. Current analyses usually adopt a modified universal relation, $\bphi = 2\delta_c(b_1-p)$, with $p$ calibrated to simulations. This prescription therefore relies on the accuracy of the simulated galaxy-formation model and ignores the uncertainty on $\bphi$, potentially biasing $\fnl$ constraints.}
   {We investigate whether $\bphi$ can be constrained from small-scale galaxy clustering using a flexible empirical galaxy-formation model.}
   {We fit an extended Subhalo Abundance Matching (SHAMe) model to small-scale clustering measurements and combine it with dark-matter-only separate-universe simulations to infer a posterior on $\bphi$. We validate this approach using a set of IllustrisTNG hydrodynamical separate-universe simulations.}
  {Without any calibration on hydrodynamical separate-universe simulations, SHAMe predicts $b_\phi$ with a median error of 10\%, compared with 38\% for a fixed halo occupation distribution (HOD) that neglects the response of the occupation to $\phi$, and 23\% for the universality relation. The residual error in SHAMe is dominated by central galaxies near the selection threshold and persists when the SHAMe stellar-mass assignment is replaced by the true TNG one, conditioned on the same subhalo properties.}
  {Small-scale clustering combined with a subhalo-based empirical model predicts most of the response of galaxy abundance to local PNG. The remainder depends on galaxy-formation physics and should enter $b_\phi$ priors as a theoretical uncertainty. This provides a route to data-informed priors on $b_\phi$, and on bias parameters more generally, without relying on externally calibrated relations.}


   \maketitle
   \nolinenumbers

\section{Introduction}

The unprecedented volume and number of galaxies probed by current and forthcoming large-scale structure surveys, such as the Dark Energy Spectroscopic Instrument~\citep[DESI,][]{desi_collaboration_2016, desi_data_2026}, Euclid~\citep{euclid_overview_2025}, LSST~\citep{lsst_2019}, or the Nancy Grace Roman Space Telescope, provide increasingly precise measurements of the clustering of galaxies. Extracting cosmological information from these data requires an accurate modeling of the galaxy density field over a broad range of scales. In particular, a central challenge is the connection between the observed galaxy population and the underlying matter distribution, which is governed by complex galaxy-formation processes that are difficult to model from first principles.

Several approaches have been developed in the literature to model this connection. On large scales, the relation between galaxies and dark matter, together with the gravitational evolution of the matter density and velocity fields, can be described perturbatively. Significant progress in this direction over the past decade has led to the development of the effective field theory of large-scale structure \citep[EFTofLSS, see e.g.][]{carrasco_effective_2012, baumann_cosmological_2012, senatore_ir-resummed_2015}. In this framework, the effects of unresolved small-scale physics, such as galaxy formation processes, are encoded in a set of free parameters, which are marginalized over when inferring cosmological parameters from galaxy-clustering measurements. This approach provides a robust and computationally efficient description of large-scale clustering without requiring a detailed model of galaxy formation. However, the generality of the bias expansion also means that information about the galaxy-matter connection is absorbed into nuisance parameters, which can be degenerate with cosmological parameters.

Modelling based on numerical simulations provides a complementary approach. $N$-body simulations allow us to go beyond the perturbative treatment and accurately describe the non-linear evolution of the dark matter distribution. On top of the dark matter distribution, empirical models are usually employed as a proxy for galaxy formation to populate $N$-body simulations with galaxies. On the one hand, halo occupation distribution \citep[HOD, ][]{peacock_halo_2000, seljak_analytic_2000, berlind_halo_2002, zheng_galaxy_2007} models specify the probability distribution of the number of galaxies in a dark matter halo conditioned on the halo mass. More recent implementations include additional dependence on secondary properties of the host halo~\citep[see e.g.][]{yuan_evidence_2021, hadzhiyska_millenniumtng_2023, rocher_desi_2023}. 

On the other hand, the subhalo abundance matching (SHAM) technique proposes to directly link galaxies to subhalos, by associating the relevant galaxy property (e.g. its stellar mass, luminosity, or star formation rate) to a subhalo property. In particular, the extended SHAM (hereafter SHAMe) model presented in \citet{contreras_flexible_2021b} introduces additional flexibility in this mapping and has been shown to accurately fit the clustering of galaxies in hydrodynamical simulations and observations down to separations of $\sim 0.1 \, \Mpch$~\citep{contreras_milleniumtng_2023, contreras_consistent_2023b, contreras_validating_2024}. Such models therefore provide a way of exploiting small-scale clustering information that is discarded in a purely perturbative treatment. Their predictions, however, remain conditional on the assumed galaxy-formation prescription and on the distribution of its parameters. 

In cosmological analyses of galaxy clustering, perturbative modelling has so far been the main approach. Simulation-based analyses have also been carried out, in which the parameters of an HOD~\citep[see e.g.][]{kobayashi_full-shape_2022, yuan_stringent_2022} or SHAM~\citep[e.g.][]{mahony_cosmological_2026, ortega-martinez_cosmological_2026} model are varied and marginalized over instead of EFT bias parameters. More recently, EFT and simulation-based approaches have been combined by using ensembles of simulated galaxy catalogues are used to build priors on the EFT bias parameters \citep{ivanov_full-shape_2024, ivanov_full-shape_2025}. This hybrid strategy keeps the computational efficiency and flexibility of perturbative modelling, while exploiting the information contained in nonlinear galaxy clustering to constrain the bias parameters. Informative priors also reduce projection effects arising from poorly constrained bias parameters that are degenerate with cosmological parameters. However, in existing approaches, these priors are set by galaxy formation models used to generated the simulated catalogues, not by observations.

In this work, we propose a data-driven implementation of this strategy. We use the observed small-scale galaxy clustering to constrain the parameters of an empirical galaxy-formation model, and then use simulations populated with this model to infer the corresponding large-scale bias parameters. Several estimators can be used for this second step, such as the probabilistic bias estimator of~\citet{stucker_probabilistic_2025}, or the separate-universe approach~\citep[e.g.][]{wagner_separate_2015, barreira_galaxy_2020} which we use here (section~\ref{sec:separate_universe}). In this way, information from nonlinear scales is propagated into observation-based priors for large-scale cosmological analyses. 

We illustrate this idea with the study of local-type primordial non-Gaussianity (PNG), and the associated bias parameter $b_\phi$. A detection of $\fnl \neq 0$ would rule out all single-field models of inflation, which predict a vanishing squeezed limit bispectrum through the consistency relations \citep{maldacena_non-gaussian_2003, creminelli_on_2003}. Local PNG is characterized by a primordial bispectrum that couples large and small-scale perturbations. In particular, local PNG induces a distinctive scale-dependent correction to the galaxy bias, which leaves a characteristic signature in the galaxy power spectrum (as well as bispectrum) on large scales.  The amplitude of this correction depends not only on the PNG amplitude $\fnl$, but also on the response of the galaxy abundance to a long-wavelength primordial potential perturbation, quantified by the bias parameter $\bphi$.

The tightest bound to date comes from the bispectrum of temperature fluctuations measured in the cosmic microwave background: $f_{\rm NL}^{\rm loc} = -0.9 \pm 5.1$ \citep{planck_2018_constraints}. The most precise constraint on $\fnl$ from the large-scale galaxy power spectrum comes from the DESI survey~\citep{chaussidon_constraining_2025}, with $\fnl = -3.6^{+9.0}_{-9.1}$. A recent reanalysis including the galaxy bispectrum tightens this constraint to $\fnl = -0.1 \pm 7.4$, and to $\fnl = 0.0 \pm 4.1$ when combining with the CMB~\citep{chudaykin_reanalizing_2026}. The uncertainty on $\fnl$ will further drop with the completion of the DESI survey, Euclid~\citep{euclid_testing_2026}, and SPHEREx~\citep{dore_cosmology_2014} that is expected to achieve $\sigma_{\fnl} \sim 1$.

However, the galaxy power spectrum constraints the product $\bphi\,\fnl$. Therefore, uncertainty in $b_\phi$ directly propagates into the inference of $\fnl$. While including the galaxy bispectrum helps break this degeneracy, informative priors are still required to obtain meaningful constraints~\citep{euclid_testing_2026}. A commonly adopted prescription for $\bphi$ assumes a universal halo mass function, which results in a direct relation between $\bphi$ and the linear galaxy bias $b_1$: $b_\phi = 2\delta_c(b_1-1)$, where $\delta_c \simeq 1.686$~\citep{dalal_imprints_2008, matarrese_effect_2008, slosar_constraints_2008}. But this relation is not generally valid for galaxies. For example, separate-universe simulations using the IllustrisTNG galaxy formation model show that stellar-mass-selected galaxies are better described by $b_\phi = 2\delta_c\left(b_1 - p\right)$ with $p \in [0.4, 0.7]$~\citep{barreira_galaxy_2020}. Additionally, \citet{slosar_constraints_2008} showed that $p = 1.6$ provides a more accurate description for host halos that have recently undergone a major merger. More generally, the relation between $b_\phi$ and $b_1$ depends on how the galaxy selection criterion responds to long-wavelength perturbations, and can therefore depend on galaxy properties and formation history. Recent PNG analyses based on the galaxy power spectrum thus adopt different prescriptions for different types of galaxies~\citep{chaussidon_constraining_2025}.

These limitations motivate an approach in which $b_\phi$ is constrained from observations. The possibility of obtaining data-informed priors on $b_\phi$ has been explored in a number of recent works. For example, \citet{sullivan_learning_2023} use a machine-learning approach trained on hydrodynamical simulations to predict $\bphi$ as a function of observable galaxy properties, while \citet{hadzhiyska_refining_2025} link $\bphi$ to Lagrangian bias parameters. Both methods rely on calibrations to simulations. Alternatively, \citet{sullivan_local_2025, dalal_estimating_2026} propose to infer $\bphi$ directly from the observed redshift evolution of tracer number counts, although in practice this signal can be difficult to disentangle from other evolving effects, such as the survey selection function. \citet{moore_informative_2026} use a semi-analytic model of galaxy formation to derive priors on $\bphi$ conditioned on observables such as the stellar mass function (SMF) and the stellar-to-halo mass relation (SHMR). \citet{perez_impact_2026} showed that variations of the stellar and AGN feedback parameters of a semi-analytic model span the range of $b_\phi(b_1)$ relations previously measured in hydrodynamical simulations and other semi-analytic models. More recently, \citet{yu_how_2026} derived $b_\phi$ priors by propagating the posterior of an HOD model fitted to the small-scale clustering of the DESI Early Data Release. However, standard HOD models assume that the occupation of haloes at fixed mass does not respond to long-wavelength perturbations, an approximation that \citet{voivodic_response_2021} showed to be inaccurate.

In this work, we follow a similar strategy to \citet{yu_how_2026} but differ in two aspects. First, we use a subhalo-based empirical model, SHAMe, in which the galaxy population responds to long-wavelength perturbations through the response of the subhalo population. Second, we validate the full procedure against hydrodynamical separate-universe simulations in which the true $b_\phi$ is known. 
The remainder of this paper is organized as follows. Section~\ref{sec:separate_universe} presents the separate-universe formalism and the simulations used in this work. In section~\ref{sec:method}, we introduce our methodology to extract a posterior on $\bphi$ from observed small-scale clustering. Section~\ref{sec:shame_inference} describes the inference of galaxy-formation parameters. Results on $\bphi$ are presented in section~\ref{sec:results}. We summarize and conclude in section~\ref{sec:conclusion}.

\section{Separate-universe simulations}\label{sec:separate_universe}

\subsection{Bias parameters and the peak-background split}

On sufficiently large scales, the galaxy density contrast can be written as:

\begin{equation}
    \delta_g (\vx, z) = \sum_{\op} b_{\op}(z)\,\op(\vx, z) + \epsilon(\vx, z)
\label{eq:bias_expansion}
\end{equation}

\noindent where $\op$ denotes operators constructed from the underlying matter field, including the tidal field, $b_{\op}$ are the bias parameters that quantify the response of the galaxy density to the corresponding operators $\op(\vx, z)$, and $\epsilon$ accounts for stochastic contributions. The bias parameters $b_{\op}(z)$ encode the impact of galaxy-formation physics on the relation between the galaxy and matter density fields.

In the peak-background split (PBS) formalism~\citep{kaiser_on_1984, cole_biased_1989, desjacques_large-scale_2018}, a sufficiently long-wavelength perturbation can be considered locally constant over the scales relevant for galaxy formation and be interpreted as a modification of the background cosmology. This allows one to separate long-wavelength perturbations from the short-scale fluctuations involved in tracer formation. As a result, the response of a tracer population to a long-wavelength fluctuation can be measured by comparing otherwise identical universes with a different background parameter.

\subsection{Local PNG and scale-dependent bias}

In the presence of local PNG, the primordial gravitational potential $\phi$ reads:
\begin{equation}
\label{eq:phi}
    \phi(\vx) = \phi_{\rm G}(\vx) + \fnl \left( \phi_{\rm G}(\vx)^2 - \avg{\phi_{\rm G}^2}  \right)
\end{equation}

\noindent where $\avg{...}$ denotes ensemble average. On sufficiently large scales, equation~\eqref{eq:bias_expansion} reduces at linear order ro:
\begin{equation}
    \delta_g (\vx, z) = b_1(z) \delta_m (\vx, z) + \bphi(z)\, \fnl\, \phi(\vx) + \epsilon(\vx, z).
\end{equation}

\noindent At the level of the galaxy power spectrum, this leads to the characteristic scale-dependent galaxy bias:
\begin{equation}
    P_{gg}(k, z) = \left[ b_1 + \frac{\bphi \fnl}{\mathcal{M}(k, z)} \right]^2 P_{mm}(k, z) + P_{\epsilon\epsilon}(k, z),
\label{eq:scale_dependent_bias}
\end{equation}

\noindent where $\mathcal{M}(k,z) = 2 c^2 k^2 T_m(k) D(z) / (3\Omo H_0^2)$, with $T_m(k)$ the matter transfer function, normalised to unity on large scales, and $D(z)$ the linear growth factor, normalised to $(1+z)^{-1}$ during matter domination. Therefore, local PNG is imprinted in the galaxy power spectrum on large scales. 

In the presence of local PNG, the long-wavelength primordial potential locally modulates the amplitude of small-scale primordial fluctuations. Splitting $\phi_G = \phi_L + \phi_s$ in equation~\ref{eq:phi} into long ($\phi_L$) and short ($\phi_s$) contributions, the short-wavelength potential $\phi_s$ becomes at first order in $\fnl$:
\begin{equation}
    \phi_s \xrightarrow{} \phi_s (1 + 2\fnl \phi_L)
\end{equation}

\noindent The long mode locally rescales the amplitude of small-scale fluctuations by $\Delta \sig = 2f_{\rm NL}^{\rm loc}\phi_L$\footnote{We denote $\Delta \sig$ the \textit{fractional} variation of $\sig$: $\sig \xrightarrow{} \sig (1 + \Delta \sig)$.}. Within the PBS framework, the response of the galaxy number density to this modulation defines $\bphi$:
\begin{equation}
    \bphi (z) = \frac{\mathrm{d} \ln \bar{n}_g (z)}{\mathrm{d} \left( \fnl \phi_L \right)} = 2 \left. \frac{\mathrm{d} \ln \bar{n}_g (z)}{\mathrm{d} (\Delta \sigma_8) } \right|_{\Delta \sig = 0}
\label{eq:bphi_su}
\end{equation}

\noindent Therefore, varying the background value of $\sig$ while keeping the other cosmological parameters fixed allows us to measure the response of the galaxy abundance to the local modulation induced by PNG. In practice, we estimate this derivative using three simulations with different values of $\sig$ around a fiducial cosmology with $\Delta \sig = \pm 0.025$.

\begin{figure*}
\centering    
\includegraphics[width=\hsize]{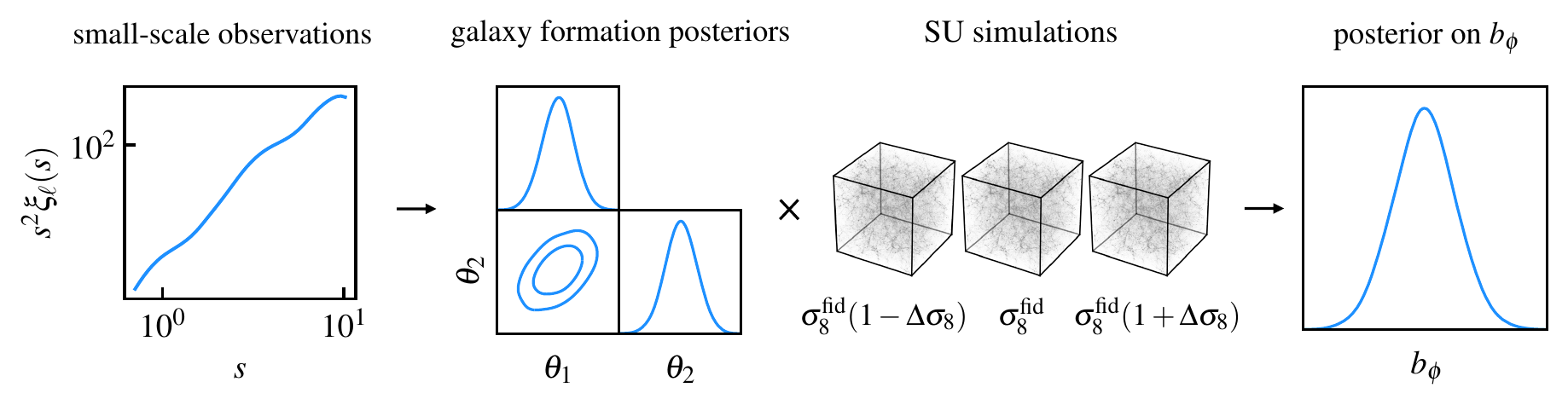}
  \caption{Schematic view of our method. The small-scale clustering measurements of the fiducial TNG galaxy sample are used to constrain the parameters of the SHAMe galaxy-formation model through MCMC sampling. The resulting samples of SHAMe parameters are then used to populate three dark-matter-only separate-universe simulations, yielding a $\bphi$ posterior that captures uncertainties in galaxy formation parameters.}
\label{fig:scheme_pipeline}
\end{figure*}

\subsection{TNG separate-universe simulations}\label{sec:tng_su}

To validate our approach, we use the hydrodynamical separate-universe simulations presented in \citet{barreira_galaxy_2020}, that were created with the IllustrisTNG  model~\citep{weinberger_simulating_2017, pillepich_simulating_2018, nelson_illustristng_2019}. These simulations were generated in a cubic box with size $L = 205 \, \Mpch$, and contain $N = 1250^3$ dark matter particles and initial gas cells. The fiducial cosmology of these simulations is the best-fit (TT,TE,EE+lowP+lensing+ext) cosmology of \citet{planck_2015_constraints}, with cosmological parameters summarized in table~\ref{tab:TNG_SU_cosmology}. The two separate universes --- that vary $\sig$ around the fiducial value --- have the same initial conditions, resolution and parameters as the fiducial simulation. In the following, we will refer to these simulations as TNG-SU. The TNG-SU simulations are available in five redshift snapshots: $z = 0, 0.5, 1, 2, 3$. In this work, we will focus on the snapshots $z = 0, 1, 2$, as our SHAMe emulator covers redshifts in the range $z \in [0, 2]$ (section~\ref{sec:emulator}). 

To create mock galaxy catalogues from the TNG-SU simulations, we select galaxies according to their stellar mass $\mstell$. We define galaxies as subhalos found by the SUBFIND algorithm~\citep{springel_populating_2001} within halos which are themselves identified as Friends-of-Friends (FoF) structures~\citep{davis_evolution_1985}. We define the stellar mass as the sum of the masses of all star particles within twice the stellar half mass radius of the subhalo. We measure $\bphi$ for four number densities, $\log \nbar = \{ -4,\, -3.5,\, -3,\, -2.5 \}$. We select galaxies by fixing the stellar-mass threshold corresponding to the target number density in the fiducial universe. The same threshold is applied to all three TNG-SU simulations, so that the resulting variation in the number of galaxies $N_g$ across the three simulations is the response to $\sig$. 

To estimate $\bphi$, we fit a linear function to the three $\left(\Delta\sig, \ln  N_g \right)$ points:
\begin{equation}
    \ln N_g(\mstellcut, \Delta\sig) = A + \frac{\bphi(\mstellcut) \Delta\sig}{2}.
\label{eq:bphi_line_fit}
\end{equation}

\noindent We estimate errors on $N_g$ using jackknife resampling: we split the three separate-universe simulations in $N_{\rm JN} = 125$ sub-boxes, remove the same subbox from the three separate universes and compute the resulting counts (scaled by the volume fraction). This jackknife estimate of the error on $b_\phi$ is similar or smaller than the finite-difference error estimate of \citet{barreira_galaxy_2020}.

These measurements will be our reference values of $b_\phi$. In the following, we aim to test whether these values can be recovered using dark-matter-only $N$-body simulations and an empirical galaxy formation model calibrated on the small-scale clustering of the TNG fiducial simulation. We treat the latter as our ``observed'' data vector, mimicking the situation in which the underlying galaxy formation physics is unknown and only galaxy clustering measurements are available.

\begin{table}
\centering
\begin{tabular}{cccccc}
 & $\Om$ & $\Ob$ & $h$ & $\sig$ & $n_s$  \\
\hline
fiducial & 0.3089 &  0.0486 & 0.6774 & 0.8159 & 0.9667\\
low $\sig$ & - & - & - & 0.7953 & - \\
high $\sig$ & - & - & - & 0.8360 & - \\
\hline
\end{tabular}
\caption{Cosmology of the TNG-SU simulations.}
\label{tab:TNG_SU_cosmology}
\end{table}

\subsection{TNG-mimic separate-universe simulations}\label{sec:tng_mimic}

While hydrodynamical simulations allow us to  directly model galaxy formation by simultaneously resolving the evolution of dark matter and baryons, their computational cost prevents us from running large suites of such simulations with different galaxy formation models and cosmologies. 

To construct a realistic setting in which the galaxy formation model is not known a priori, we complement the TNG-SU simulations with a set of dark-matter-only simulations that reproduce the dark matter and subhalo populations of the TNG-SU suite. We construct these simulations with the gravity-only code \texttt{LGADGET-3}~\citep{angulo_bacco_2021}, that builds on \texttt{GADGET-2}~\citep{springel_cosmological_2005}. Halos are then identified on-the-fly using the FoF algorithm, and subhalos are found using a modified version of SUBFIND. The latter takes into account the past history of the subhalo, while also measuring properties that are non-local in time such as the peak maximum circular velocity $\vpeak$. Running each simulation required around 9h of computing time using two nodes (48 CPU each) on DIPC's supercomputer Hyperion~\footnote{\url{https://scc.dipc.org/docs/systems/hyperion/}}. We refer to these as the TNG-mimic-SU simulations. 

The TNG-mimic-SU are constructed with the same cosmological parameters and initial phases as their TNG counterparts, in the same volume, with a resolution of $N = 625^3$ dark matter particles, i.e.\ a particle mass of $\approx 3\times10^{9}\,h^{-1}{\rm M}_\odot$. This is eight times coarser than the dark matter in TNG-SU, but comparable to the mass resolution of the simulations used to train the SHAMe emulator (section~\ref{sec:emulator}), which ensures that the SHAMe parameters inferred with the emulator apply to the TNG-mimic-SU subhalo catalogues. Unlike the TNG-SU simulations, the TNG-mimic-SU do not contain galaxies or any prescription for galaxy formation. The role of the TNG-mimic-SU simulations is to provide the response of the subhalo population to a change in the amplitude of matter fluctuations, which can then be combined with different galaxy--subhalo connections. In the next section, we detail our procedure to measure $\bphi$ from these simulations.

\section{Methodology}\label{sec:method}

\subsection{General idea}\label{sec:overview}

Our goal is to infer $b_\phi$ (or, more generally, galaxy bias parameters $b_{\op}$) for a given galaxy sample using its small-scale clustering. We use these measurements to constrain the parameters of an empirical galaxy-formation model, then propagate their posterior uncertainty to $b_\phi$ through separate-universe simulations. Our approach is summarized in figure~\ref{fig:scheme_pipeline}.

Let $X$ denote a small-scale observable that probes the galaxy-matter connection, e.g. the small-scale galaxy correlation function, and let $\theta = (\thetac, \thetagf)$ denote the parameters describing the cosmological and galaxy-formation models. Observations provide a posterior:

\begin{equation}
    \probdens{\theta\, | \, X} \propto \probdens{X\,|\,\theta} \probdens{\theta}.
\end{equation}

\noindent Simulations implementing this galaxy formation model can then be used to estimate bias parameters $b_{\op}$. Marginalizing over the posterior uncertainty in $\theta$ gives:
\begin{equation}
    \probdens{b_{\op}\,|\,X} = \int d\theta \probdens{b_{\op}\,|\,\theta} \probdens{\theta\,|\,X}.
\label{eq:bias_posterior}
\end{equation}

\noindent Here, $\probdens{\theta\, |\, X}$ is constrained by small-scale observations and the adopted galaxy-formation model, while $\probdens{b_{\op}\, |\, \theta}$ is determined from simulations. The resulting posterior on $b_{\op}$ can be used as an informative prior in subsequent large-scale cosmological analyses. This framework is not restricted to a particular bias parameter or galaxy-formation prescription.

In this work, we apply this approach to infer $b_\phi$ using the small-scale redshift-space correlation function $\xi$ of the fiducial TNG galaxy sample. Its configuration-space representation allows us to isolate the scales most relevant to the galaxy-subhalo connection. For each galaxy selection, we fit the measured $\xi$ multipoles with an emulator of the SHAMe galaxy-formation model (section~\ref{sec:emulator}), obtaining the posterior:
\begin{equation}
\probdens{\thetashame\,|\,\xi} \propto \probdens{\xi\,|\,\thetashame} \probdens{\thetashame}.
\end{equation}

\noindent For each posterior sample $\theta_{\rm SHAMe}$, we then populate the three TNG-mimic-SU simulations using the same SHAMe parameters in each. We refer to these realizations as SHAMe-SU simulations. Their predictions for $b_\phi$ can be marginalized over the SHAMe posterior:
\begin{equation}
    \probdens{\bphi\, |\, \xi} = \int d\thetashame \probdens{\bphi\, |\, \thetashame} \probdens{\thetashame\, |\, \xi}.
\label{eq:bphi_posterior}
\end{equation}

In general, $b_\phi$ can also depend on cosmological parameters. Our framework can be extended to account for this dependence by including cosmological parameters in $\theta$ and generating a suite of separate-universe $N$-body simulations spanning the relevant parameter space.  In this work, for simplicity, we fix the cosmological parameters to their fiducial TNG values and vary only the galaxy-formation parameters. 

\subsection{Separating the galaxy--subhalo connection from the separate-universe response}

The approach above assumes that the response of the galaxy abundance to $\sig$ enters only through the subhalo population. If $\xsub$ denotes the subhalo properties that set the galaxy stellar mass in the model, we assume:

\begin{equation}
\begin{split}    
    \probdens{\mstell\, |\, \xsub, \thetashame, \sig} 
    &= \probdens{\mstell\, |\, \xsub, \thetashame}.
\end{split}
\label{eq:prob_mstell_sigma8}
\end{equation}

Marginalizing over subhalo properties $\xsub$, we can write:

\begin{equation}
\begin{split}    
    \probdens{\mstell\, |\, \thetashame, \sig} = \int d\xsub \probdens{\mstell\, |\, \xsub, \thetashame} \probdens{\xsub\, |\, \sig}.
\end{split}
\label{eq:prob_mstell_factorized}
\end{equation}

Under this assumption, the $\sig$ dependence of the abundance of galaxies above a $\mstell$ threshold arises completely from changes in the subhalo population $\probdens{\xsub\,|\,\sig}$. We can measure this response directly from dark-matter-only separate-universe simulations. The galaxy--subhalo relation, $\probdens{\mstell\,|\,\xsub,\thetashame}$, is then determined by the SHAMe model fitted to observed small-scale clustering in the fiducial universe. 

Therefore, our approach has two assumptions: i) $\thetashame$ are the same in all three separate universe universes, so $\probdens{\theta_{\rm shame}\,|\,\xi}$ from the fiducial universe also applies to the other two; and ii) the set $\xsub$ used by SHAMe captures the full $ \sig$ dependence of $\mstell$. These assumptions would fail, for instance, if the stellar mass of central galaxies at fixed $V_{\rm peak}$ depended on halo formation time, and formation time responded to $\sigma_8$ differently from $V_{\rm peak}$.

We test these two assumptions separately. First, in section~\ref{sec:shame_inference}, we fit the three TNG-SU simulations and check whether they give consistent SHAMe parameters. This test probes galaxy clustering at fixed number density, so it is sensitive to our first assumption but not to the abundance response that determines $b_\phi$. Therefore, in section~\ref{sec:results} we test the second assumption directly. We match central subhaloes between the TNG-SU and TNG-mimic-SU simulations, measure $P(M_\star > M_\star^{\rm cut}\,|\,\mathcal{S})$ in the fiducial TNG-SU, and apply it to the three TNG-mimic-SU simulations. This provides an estimate of $b_\phi$ that is free of any error from the SHAMe parametrisation and fit, so any remaining difference from the true $\bphi$ isolates the errors of the second assumption.

In the next subsection, we describe our practical estimator for $N_g$.

\subsection{Sampling $\bphi$ posterior with the SHAMe-SU mocks}\label{sec:shame_bphi_estimator}

We draw $N_s=100$ samples $\{\theta_i\}$ from the SHAMe posterior $\probdens{\thetashame\, |\, \xi}$ (section~\ref{sec:shame_inference}). Each sample also employs an independent SHAMe random seed $s_i$ to account for the stochasticity of the SHAMe algorithm.

SHAMe cannot be used directly to assign masses to the SU simulations, since it abundance matches to an input SMF. Using a fiducial SMF in the other SU simulations would keep $N_g$ fixed and the response would be zero, $b_\phi=0$ by construction. Additionally, in a real application, the separate-universe SMF is unknown, and thus the method has to infer how $\sig$ changes it. For these reasons, we use SHAMe only in the fiducial SU simulations, then learn the galaxy-subhalo relation, and apply it to the other two SU simulations at fixed subhalo properties.

In practice, for each posterior sample $\theta_i$, we populate the fiducial TNG-mimic-SU simulation with SHAMe galaxies using the fiducial SMF. We then train a Random Forest (RF) classifier\footnote{As implemented in \texttt{scikit-learn} \citep{pedregosa_scikit-learn_2011}.} to predict whether a subhalo hosts a galaxy above the stellar-mass threshold, given its properties $\xsub$: $\prob{\mstell > \mstellcut\, |\, \xsub, \theta_i, s_i}$. As inputs, we use all properties that enter the SHAMe algorithm (section~\ref{sec:SHAMe}): $V_{\rm peak}$, time since accretion, time since becoming an orphan, and per-object bias. Applying the RF to a subhalo catalogue of a SU simulation gives an estimate of the galaxy abundance in that universe  $\hat{N}_g(\mstellcut,\theta_i,s_i,\sig)$ using equation~\eqref{eq:prob_mstell_factorized}. We evaluate this expression through Monte Carlo integration:

\begin{equation}
    \hat{N}_g(\mstell > \mstellcut, \theta_i, s_i, \sig) = \sum_{\rm subhalos\, \in\, SU(\sig)} \prob{\left. \mstell > \mstellcut \right| \xsub, \theta_i, s_i},
\label{eq:ng_estimator}
\end{equation}

\noindent where $\prob{\mstell > \mstellcut \,| \, \xsub, \theta_i, s_i} = \int_{> \mstellcut} d\mstell \probdens{ \mstell \,|\, \xsub, \theta_i, s_i}$. 

We verified that equation~\eqref{eq:ng_estimator} reproduces $N_g$ in the fiducial SHAMe-SU mock to within 0.6\% (which is at most $5\%$ of $N_g^{\rm SU} - N_g^{\rm fid}$).

Finally, for each $\theta_i$ we measure $b_\phi$ using the estimator of equation~\eqref{eq:bphi_line_fit}. The resulting set of $b_\phi$ estimates provides a Monte Carlo realization of the marginalized posterior $\probdens{b_\phi(\mstellcut)|\xi}$.

\section{Constraints on galaxy formation from small-scale clustering}\label{sec:shame_inference}

We first test the first part of the assumption in equation~\ref{eq:prob_mstell_sigma8} by fitting the three TNG-SU simulations independently with the SHAMe model and checking whether the inferred parameter posteriors are consistent across separate universes.

\subsection{Observable}\label{sec:observable}

\begin{figure*}
    \centering
    \includegraphics[width=\hsize]{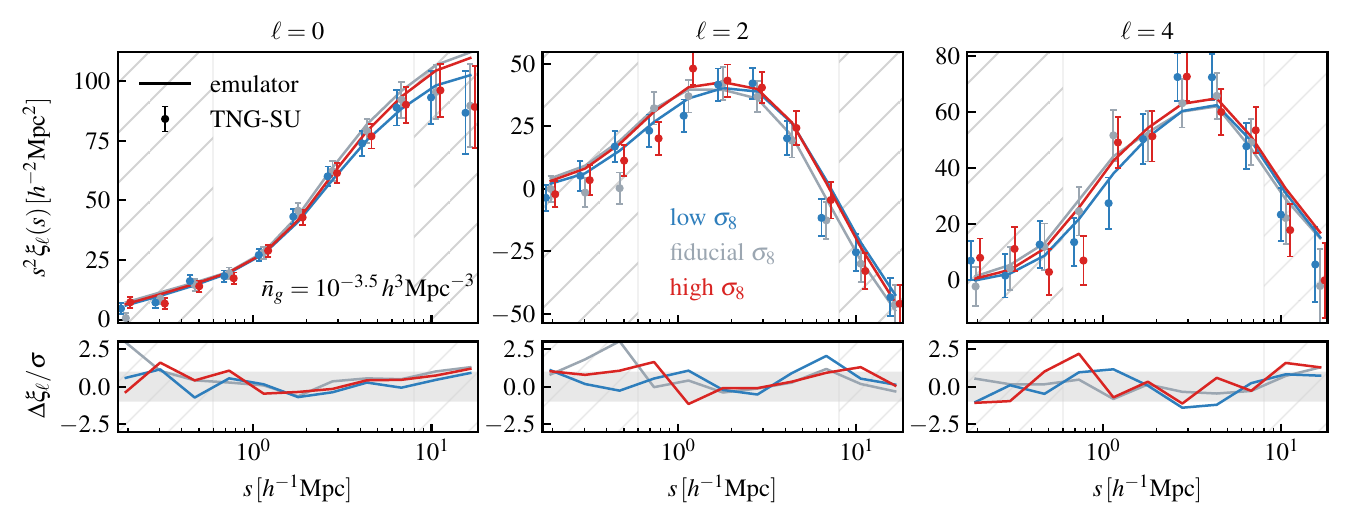}
    \caption{TNG-SU correlation function multipoles for galaxies selected by stellar mass at a density $\nbar = 10^{-3.5} \, h^3 \mathrm{Mpc}^{-3}$, at $z = 0$. Blue, grey and red points show the correlation function for the low, fiducial and high $\sig$, respectively. Low and high amplitudes are shown with a small $s$-axis offset for clarity. Solid lines show the best-fit emulator model for each amplitude. Hatched regions indicate scales that were not used in the fit.}
    \label{fig:bestfit_example}
\end{figure*}

Our observable to constrain galaxy formation parameters is the set of small-scale redshift-space correlation function multipoles $\xi_{\ell}$, with $\ell = (0, 2, 4)$. We apply Alcock-Paczynski distortion with the same cosmology as the emulator and average our measurements over the three axes of the box. We measure the correlation function multipoles in 15 log-spaced bins in the range $[0.1, 78] \, \Mpch$, integrating over 40 $\mu$-bins in $[0, 1]$. These bins correspond to those used to generate the training set for the emulator described in section~\ref{sec:emulator}. 
We restrict the fit to scales larger than $\smin \sim 0.5 \,\Mpch$, where the model has been validated, and smaller than $\smax \sim 10 \, \Mpch$ to limit finite-box-size effects (appendix \ref{sec:boxsize}). We show the TNG-SU correlation function multipoles for $\bar{n}_g = 10^{-3.5} \, \nunits$ at $z = 0$ in figure~\ref{fig:bestfit_example}, with hatched regions indicating separations excluded in the fit. 

\subsection{Extended SHAM galaxy formation model}\label{sec:SHAMe}

The most basic SHAM implementation consists of ranking subhalos in a simulation by a given property and then associating a galaxy to each subhalo in decreasing order of this property, with stellar mass (or any property relevant to the galaxy selection) determined by the SMF of the observed galaxy population. Additionally, SHAM implementations typically allow some scatter in the relationship between the subhalo property and the galaxy property. We note that, for samples defined by a target number density, the predicted clustering is largely independent of the shape of the SMF. We have checked that the impact of a misspecified SMF on $b_\phi$ is less than 15\%, and within the scatter allowed by SHAMe parameters (see appendix~\ref{sec:misspecified_smf}).

In this work, we use as the relevant subhalo property the peak value of the maximum circular velocity that a subhalo has reached throughout its history up to the redshift of interest, $\vpeak$. The maximum circular velocity $\vmax = \max_r \left[ \sqrt{G M(<r)/r)} \right]$ serves as a proxy for the depth of the gravitational potential well, which is a good indicator of the integrated gas mass available for star formation~\citep{chaves-montero_subhalo_2016}. Using the peak value of $\vmax$ also makes it relatively insensitive to tidal stripping after infall. $\vpeak$ was shown to be one of the properties that best predict the stellar mass~\citep{chaves-montero_subhalo_2016}. We consider the extended SHAM implementation of \citet{contreras_flexible_2021b, contreras_consistent_2023a}, which introduces four additional parameters to overcome some of the limitations of the SHAM algorithm, allowing us to consistently reproduce clustering and galaxy-galaxy lensing statistics in state-of-the-art hydrodynamical simulations~\citep{contreras_consistent_2023a, contreras_milleniumtng_2023, contreras_validating_2024}, as well as real observations~\citep{contreras_consistent_2023b}. The detailed SHAMe algorithm is presented in appendix~\ref{sec:SHAMe_algo}. The resulting five parameters of the SHAMe model are:

\begin{itemize}
    \item $\sigmalum$, which controls the scatter between the galaxy stellar mass and $\vpeak$;
    \item $\logtmerger$, which regulates the number of orphan\footnote{Orphans are satellite galaxies whose subhalo is no longer resolved by the simulation at the time considered (due to tidal stripping), but have not yet been completely destroyed or merged with the central structure.} galaxies. $\logtmerger$ effectively removes from the simulation subhalos that have been orphans for a long time;
    \item $\fkp$, $\fkm$, which control the level of assembly bias, in this case the dependence of the stellar mass on the local linear galaxy bias. Specifically, $\fkp$ and $\fkm$ encode the sum and the difference between the amount of galaxy assembly bias added to central and satellite galaxies, respectively~\citep[see][]{contreras_flexible_2021a};
    \item $\betalum$, which sets the time scale during which satellite galaxies survive within their host halo before losing all their stellar mass or exit the selected catalogue due to e.g. -- quenching, tidal stripping, or reddening. Satellite galaxies that have been satellites for a long time are therefore removed from the sample~\citep{contreras_consistent_2023a}.
\end{itemize}

\subsection{Emulator}\label{sec:emulator}

To model the galaxy correlation function and constrain galaxy formation parameters, we use the publicly available emulator\footnote{\url{https://github.com/hantke/GalaxyEmu}} described in \citet{mahony_cosmological_2026}, which is trained on a suite of $N$-body simulations populated with SHAMe galaxies. 

The suite of $N$-body simulations is built from five pairs of simulations, run with the same code as the TNG-mimic-SU simulations (section~\ref{sec:tng_mimic}), with a resolution of $3072^3$ particles over a $(1024 \, \Mpch)^3$ volume. The pairs of simulations have opposite initial Fourier phases to reduce the effect of cosmic variance, following the method from \citet{angulo_cosmological_2016}. These five simulations are then scaled to \num{10000} different cosmologies using the ``cosmology rescaling'' technique from \citet{angulo_one_2010, contreras_3percent_2020}. This technique allows us to modify the outputs of a dark matter simulation by displacing its particles, haloes and subhaloes to reproduce their distribution in a different cosmology, without having to re-run the $N$-body code. The simulations are generated at about 8 redshift snapshots in the range $[0, 2]$. The ranges of cosmological parameters covered by these \num{10000} simulations is reported in table~\ref{tab:emu_cosmo_params}. Two SHAMe mocks are then generated for each cosmology, with parameters spanning the ranges shown in table~\ref{tab:emu_cosmo_params}.

For each SHAMe mock, 8 galaxy samples with densities in $[10^{-4}, 10^{-2.25}] \, \nunits$ are then constructed. This leads to around 1.2 million catalogues varying the 13 parameters shown in table~\ref{tab:emu_cosmo_params}. 800 are excluded from the training set and kept for testing. For each sample, the correlation function multipoles ($\ell \in (0, 2, 4)$) are computed in 15 log-spaced bins in the range $[0.1, 78] \, \Mpch$, and 11 linearly spaced bins in the range $[85, 140] \, \Mpch$, integrating Legendre multipoles over 40 $\mu$-bins in $[0, 1]$. Galaxies are placed in redshift space in the plane parallel approximation and the multipoles are averaged over the three axes of the simulation box. Alcock-Paczynski effect was incorporated in the measurements, with fiducial cosmology equal to $Planck$ 2018 cosmology\footnote{The mean values from the \texttt{\seqsplit{base\_plikHM\_TTTEEE\_lowl\_lowE\_lensing}} posterior.}~\citep{planck_2018_constraints}.

A feed-forward neural network was trained to predict each multipole as a function of the parameters listed in table~\ref{tab:emu_cosmo_params}. Each network consists of two fully connected hidden layers, each with 200 neurons, and a rectified linear unit activation function. The neural networks were trained with the Keras front-end of the TensorFlow library. The Adam optimization algorithm was used, with a learning rate of 0.001 and a mean squared error loss function.

\begin{table}
\centering
\begin{tabular}{ccccc}
parameter & range \\
\hline
$\sig$ & $[0.65, 0.9]$ \\
$\Om$ & $[0.23, 0.4]$ \\
$\Ob$ & $[0.04, 0.06]$ \\
$h$ & $[0.6, 0.8]$ \\
$n_s$ & $[0.92, 1.01]$ \\
$M_{\nu} \, [\mathrm{eV}]$ & $[0, 0.4]$ \\
\hline
$\sigmalum$ & $[0, 1.8]$ \\
$\logtmerger$ & $[-1.5, 1.2]$ \\
$\fkp$ & $[-0.5, 0.5]$ \\
$\fkm$ & $[-0.5, 0.5]$ \\
$\betalum$ & $[0, 1]$ \\
\hline
$\mathrm{log}(\nbar \, [\nunits])$ & $[-4, -2.25]$ \\
$z$ & $[0, 2]$\\
\end{tabular}
\caption{Parameters of the emulator, with their varied ranges. The ranges of the SHAMe parameters (middle panel) are also the boundaries of the uniform priors used in inference. Other parameters are fixed to their true values.}
\label{tab:emu_cosmo_params}
\end{table}

\subsection{Covariance}

We build a covariance matrix that accounts for the different sources of errors in our inference. It includes cosmic variance, that is intrinsic to the data sample, as well as errors related to our emulator model. We estimate the sample error by populating a $(2048 \, \Mpch)^3$ $N$-body simulation with SHAMe galaxies, using the parameters that best fit the fiducial TNG simulation, and computing the covariance of multipoles measured in 1000 subboxes of $204.8\,\Mpch$, matching the TNG volume. We fit the SHAMe model directly by populating the TNG-mimic fiducial simulation with SHAMe galaxies on-the-fly, using a particle swarm optimisation (PSO) algorithm~\citep{duane_hybrid_1987} as implemented in~\citet{arico_bacco_2021}\footnote{\url{https://github.com/hantke/pso_bacco}}. The covariance used in this direct fit is based on jackknife resampling with $343$ jackknife regions. The $N$-body simulation used for covariance has a different cosmology than the TNG fiducial one, but we have checked that this has a minor impact on the variance (appendix~\ref{sec:data_covariance}).

Moreover, the finite size of our simulated data (which is around 125 times smaller than the emulator training data) may introduce finite box-size effects, due to the missing large-scale modes. In appendix~\ref{sec:boxsize}, we show that this effect is below $\sim 20\%$ of our estimated data error at the scales considered here.

The model error arises at various levels: the scaling procedure used to generate the emulator training sample, the error intrinsic to SHAMe assumptions, and the emulator interpolation error. In appendix~\ref{sec:shame_error}, we show that a direct fit of the SHAMe model applied to the TNG-mimic fiducial simulation is in good agreement with the TNG fiducial simulation, thus we ignore this error. \cite{contreras_milleniumtng_2023} also showed that this error is sub-dominant compared to scaling and emulator errors. 

We estimate the emulator error using the emulator test set (appendix~\ref{sec:emu_cov}). The scaling error is estimated from an ensemble of pairs of scaled and target simulations with varying cosmologies and SHAMe parameters (appendix~\ref{sec:scaling_cov}). Our final covariance estimate is therefore the combination of three covariance matrices:
\begin{equation}
    C_{\rm tot} = C_{\rm data} + C_{\rm emulator} + C_{\rm scaling}.
\end{equation}
Note that we use the same covariance for the fits to the three TNG-SU simulations.

\subsection{Inference of SHAMe parameters}

\begin{figure*}[ht!]
    \centering
    \includegraphics[width=\hsize]{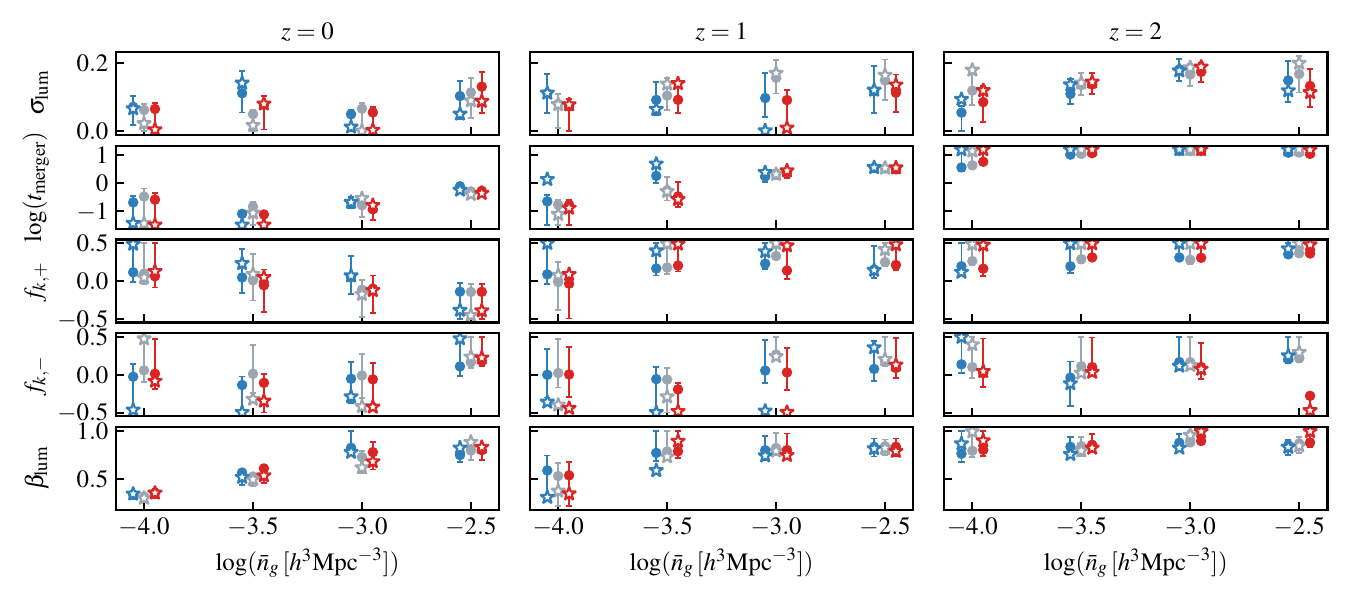}
    \caption{Constraints ($1 \sigma$ credible intervals) on SHAMe parameters from $\xi_{\ell}(s)$ measured on TNG-SU simulations. Circles show the mean of the marginalized posteriors, while stars show the maximum of the marginalized posteriors. Blue, grey and red points correspond to low, fiducial and high $\sig$ values, respectively.}
    \label{fig:shame_constraints}
\end{figure*}

To infer posteriors on SHAMe parameters, we assume a Gaussian likelihood:
\begin{equation}
\begin{split}
    \like{\theta | \xi_{\rm data}} 
    &= \exp \left\{ -\frac{1}{2}\left[ \xi_{\rm data} - \xi_{\rm emu}(\theta) \right]^T C_{\rm tot}^{-1} \left[ \xi_{\rm data} - \xi_{\rm emu}(\theta) \right] \right\},    
\end{split}
\end{equation}
with $\xi_{\rm data}$ the TNG-SU data vector (section \ref{sec:observable}) and $\xi_{\rm emu}(\theta)$ the emulator prediction. In our case, $\theta = \thetashame = (\sigmalum, \logtmerger, \fkp, \fkm, \betalum)$. We adopt uniform priors on these parameters,  with ranges equal to those covered by the emulator (table~\ref{tab:emu_cosmo_params}). Other parameters are fixed to their true values, i.e. those in table~\ref{tab:TNG_SU_cosmology} for cosmological parameters, and their corresponding $\nbar$, $z$ values for each sample.

For each galaxy sample, we sample the SHAMe posterior by using Monte Carlo Markov Chain (MCMC) as implemented in the \texttt{emcee}\footnote{\url{https://github.com/dfm/emcee}} sampler~\citep{foreman-mackey_emcee_2013}. We use \num{2000} walkers, as the emulator is more efficient when evaluated on vectorized inputs, and consider an effective sample size of 100 as convergence criterion. In figure~\ref{fig:bestfit_example}, we display the best-fit model for the three TNG-SU simulations, for $\nbar = 10^{-3.5} \, h^3 \mathrm{Mpc}^{-3}$ and $z = 0$. We see that our model is a good fit to the data in the range of scales considered here. For all samples, the best-fit $\chi^2/n_{\rm dof}$ values are between 0.2 and 1.

In practice, we will only use the fit to the TNG fiducial simulation, to mimic a realistic dataset. Therefore we check that the fiducial fit gives SHAMe posteriors similar to those of the low/high amplitude fits. Figure~\ref{fig:shame_constraints} shows the $1\sigma$ credible intervals on $\thetashame$ parameters for all densities and redshift considered in this work. We display the mean of the marginalized posterior (circles) as well as the maximum of the marginalized posterior (star), as our posteriors exhibit some prior projection effects. In addition to the constraints for the TNG-SU fiducial simulation, we show the constraints obtained for the two TNG separate universes (in red and blue, respectively). 

The constraints from the three separate universes are compatible within $1\sigma$ for most parameters and samples, the largest difference being for $\sigma_{\rm lum}$ at $z=0$ and $\nbar = 10^{-3.5}\,\nunits$. Several parameters, however, reach the boundaries of the emulator range (for instance, $\log t_{\rm merger}$ at $z=2$ and $f_{k,+}$ at $z=1$ and $2$). For these samples, the posterior would be truncated, artificially narrowing the resulting $b_\phi$ posterior that we will discuss in section~\ref{sec:results}.

\begin{figure*}
\centering    
\includegraphics[width=\hsize]{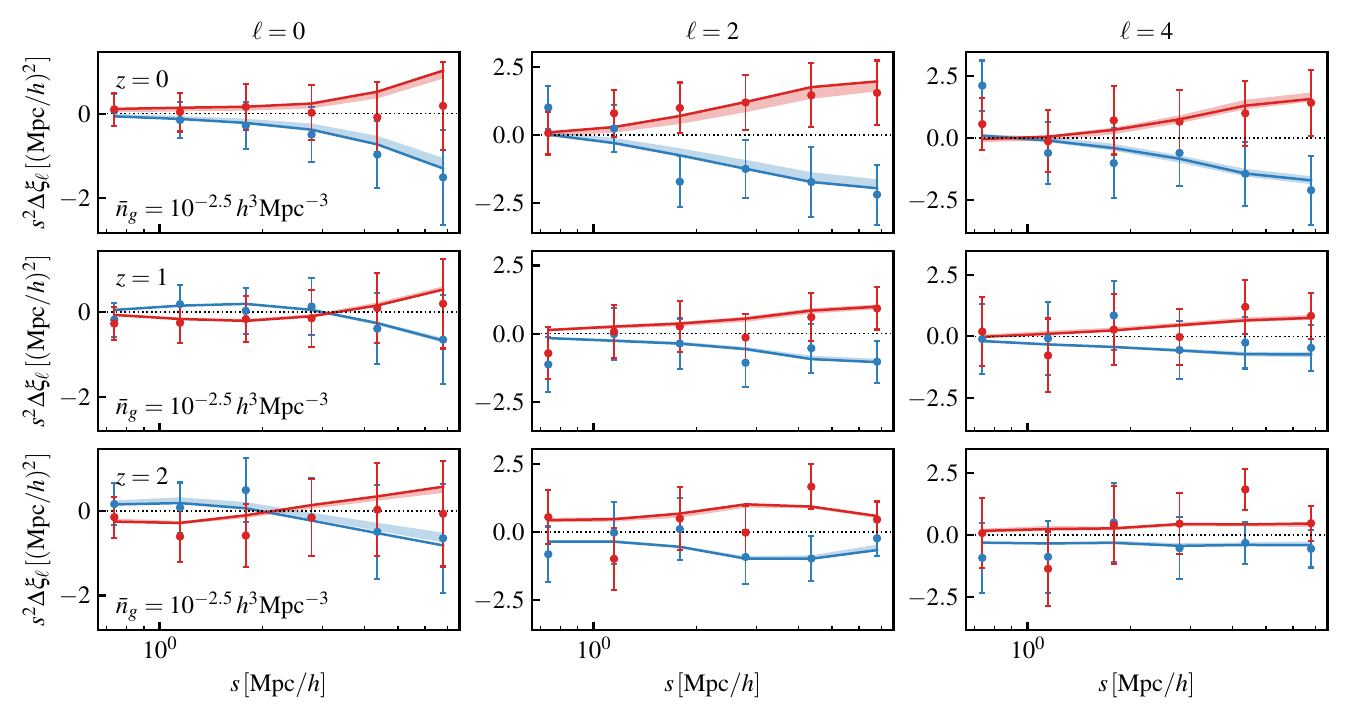}
  \caption{Difference of the correlation function multipoles $\xi_{\ell, \rm high}(s) - \xi_{\ell,\rm fid}(s)$ (red) and $\xi_{\ell, \rm low}(s) - \xi_{\ell,\rm fid}(s)$ (blue) for the highest density ($\nbar = 10^{-2.5} \, h^3 \mathrm{Mpc}^{-3}$) at $z=0$ (top row), $z=1$ (middle row) and $z=2$ (bottom row). We only show the highest density for clarity, as lower density points are noisy compared to the separate-universe differences. Points with error bars correspond to the TNG-SU simulation measurements, while solid lines show the SHAMe emulator prediction evaluated at the fiducial best-fit SHAMe parameters. Error bars are computed using the correlation matrix of the high/low and fiducial amplitudes (from jackknife realizations), with their variance scaled to match the large-box covariance matrix (which is used in inference). Emulator and scaling errors are added in quadrature to the resulting errors. Shaded areas indicate the 68\% credible interval at each point, computed from the sampled SHAMe parameters.}
\label{fig:residuals_emu}
\end{figure*}

Figure \ref{fig:residuals_emu} shows the difference between the clustering of the two separate universes and the fiducial simulation, for the highest density sample ($\bar{n}_g  = 10^{-2.5} \, h^3 \mathrm{Mpc}^{-3}$) at all redshifts. The lines show the emulator evaluated at the fiducial best-fit SHAMe parameters, for all amplitudes. We also display the per-point $68\%$ credible intervals of the fiducial SHAMe posteriors. We see that the SHAMe emulator is able to correctly predict the clustering of the separate universes, even with galaxy-formation parameters fitted to the fiducial simulation. We show residuals between the fiducial best-fit emulator prediction and the TNG-SU simulations for all densities in appendix~\ref{sec:residuals_all}.

\section{Results}\label{sec:results}

\subsection{Comparison with the true $\bphi$}\label{sec:bphi_posteriors}

\begin{figure*}
    \centering    
    \includegraphics[width=\hsize]{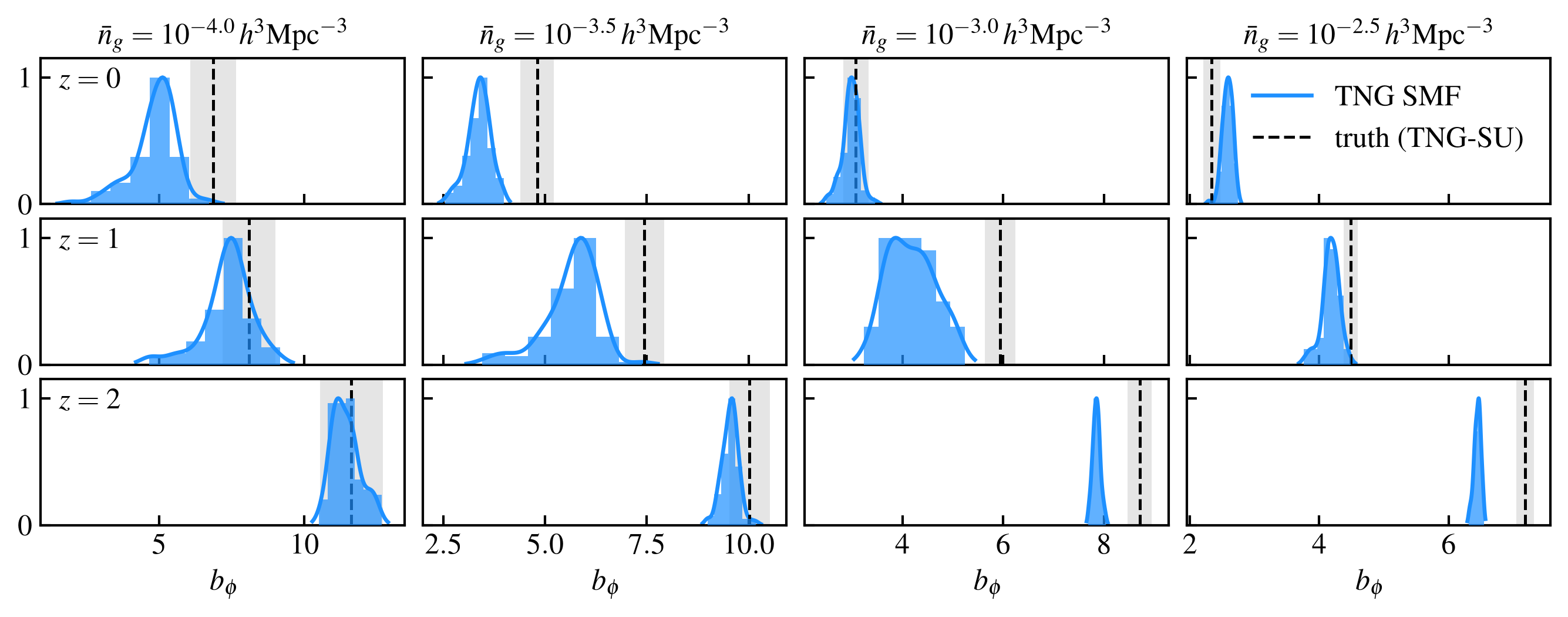}
    \caption{Measured $\bphi$ posteriors (see section~\ref{sec:shame_bphi_estimator}) compared to the prediction obtained directly from TNG-SU with the separate-universe estimator (vertical dashed line, see section~\ref{sec:tng_su}). Blue posteriors correspond to SHAMe prediction, for $N_s = 100$ samples of SHAMe parameters drawn from the fiducial posterior. Top, middle and bottom rows correspond to $z=0, 1, 2$, respectively.}
    \label{fig:bphi_measured_vs_true}
\end{figure*}

In this section, we compare the SHAMe-SU predictions for $\bphi$ (section~\ref{sec:shame_bphi_estimator}) with values measured directly in the TNG-SU (section~\ref{sec:tng_su}). Figure~\ref{fig:bphi_measured_vs_true} shows the SHAMe-SU posteriors in blue, whereas the true TNG-SU values are shown as a dashed vertical line, with shaded area equal to fit error (equation~\eqref{eq:bphi_line_fit}). 

As we do not propagate the error of each individual $\bphi$ fit into the posterior, the width of the SHAMe posteriors comes only from the uncertainty in the SHAMe parameters and from the SHAMe random seeds. Since the SHAMe-SU mocks and TNG-SU simulations share their initial conditions, any difference between the two is a systematic error in the model. 

SHAMe recovers $b_\phi$ to within $\sim10\%$ in 8 of the 12 samples, and the mean fractional error is 14\%. At $z=0$ and $z=1$, our predictions agree with the true value of $\bphi$ to within $< 3 \sigma$\footnote{We quote the significance as $|\langle b_\phi\rangle_{\rm SHAMe} - b_{\phi, \rm TNG}| / \sigma_{\rm SHAMe}$, where $\sigma_{\rm SHAMe}$ is the standard deviation of the SHAMe posterior. Since the SHAMe-SU and TNG-SU simulations share initial conditions, this quantity contains no cosmic-variance contribution.}
for most densities, but but underestimate it by around 30\% for the $\nbar = 10^{-3.5} \, \nunits$ sample at $z=0$ ($5\sigma$) and the $\nbar = 10^{-3} \, \nunits$ sample at $z=1$ ($ 3.8\sigma$). 

At $z=2$, all but the lowest density samples are underestimated by 6-10\% (up to $14 \sigma$ for the highest density sample). However, the SHAMe posteriors at high redshift are artificially narrow, because several SHAMe parameters are limited by the emulator prior (section~\ref{sec:shame_inference}, figure~\ref{fig:shame_constraints}), and thus these results should be interpreted with caution. 

This means SHAMe systematically underpredicts the response of the galaxy abundance to $\sig$. All these samples are fitted well by SHAMe. Hence, a good fit to small-scale clustering does not by itself guarantee an accurate $b_\phi$.

We now compare these predictions with commonly used $\bphi$ prescriptions, which are usually adopted without an associated uncertainty. In the next subsection, we compare them with the universality relation and with a fixed HOD. 

\subsection{Comparison with HOD and the universality relation}\label{sec:hod_universality}

We now compare our SHAMe-SU $\bphi$ predictions with two comonly used alternatives. The first is the universality relation and the second is a halo occupation distribution that does not respond to the long-wavelength perturbation, as assumed in standard HOD models. In such a model, the expectation value of the total number of galaxies above a stellar mass threshold is:
\begin{equation}
    \avg{N_g(\mstell) | \sig} = V\int dM_h \avg{N_g(\mstell) | M_h, \sigma_8} n_h(M_h | \sig),
\label{eq:prob_hod}
\end{equation}

\noindent where $n_h(M_h | \sig)$ is the halo mass function (HMF), and $\avg{N_g(\mstell) | M_h, \sigma_8}$ is the average number of galaxies with mass $\mstell$ in a halo of mass $M_h$, namely the HOD. 

We define the fixed HOD case by measuring the occupation in the fiducial TNG-SU simulation and apply it to the halo mass functions of the separate universes. Because this fixed HOD is the true occupation at the fiducial $\sigma_8$, any discrepancy with TNG-SU isolates the effect of neglecting the response of $\langle N_g(M_\star)|M_h\rangle$ to $\sigma_8$. This prediction does not include the additional uncertainty arising from fitting an HOD to clustering measurements, as done by \citet{yu_how_2026}. Therefore, it is an optimistic case for the HOD approach.

\begin{figure}[ht!]
    \centering
    \includegraphics[width=\hsize]{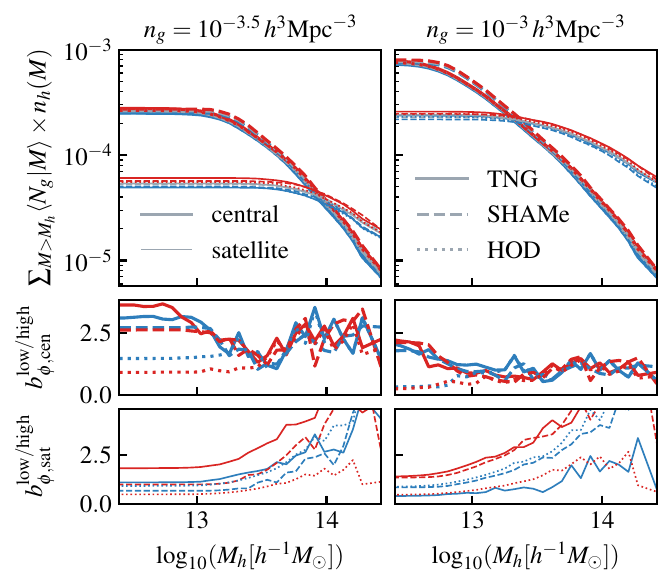}
    \caption{Cumulative HMF-weighted HOD at redshift $z=0$, split between centrals and satellites, for the TNG-SU simulations, our SHAMe-SU mocks (from the fiducial best-fit SHAMe parameters), and for a fixed-HOD prescription, i.e. assuming $\avg{N_g(\mstell) | M_h}$ has no explicit dependence on $\sigma_8$. Red and blue curves correspond to high and low $\sig$ separate universes, respectively. The second and third rows show the relative difference between the cumulative weighted HOD for the separate universes and the fiducial simulation, for centrals and satellites respectively. Both are weighted by the total number of galaxies in the fiducial simulation, so that the contribution from centrals and satellites is directly comparable.}
    \label{fig:hod_z0}
\end{figure}

In figure~\ref{fig:hod_z0}, we show the integrand of equation~\eqref{eq:prob_hod}, accumulated from high to low halo mass, for three cases: TNG-SU, the SHAMe-SU predictions at the best-fit SHAMe parameters, and the fixed HOD. We show a cumulative version of $\avg{N_g(\mstell) | M_h, \sigma_8} n_h(M_h | \sig)$ to decrease noise and better highlight the impact of these measurements on $\bphi$. We show it for two $z=0$ samples where the SHAMe error is the smallest ($\nbar = 10^{-3} \, \nunits$) and largest ($\nbar = 10^{-3.5} \, \nunits$). The bottom panels show the difference between the separate universes and the fiducial simulations for centrals and satellites separately. Each difference is normalized by the fiducial number of all galaxies and by $\Delta \sig/2$. Hence, the prediction for $b_\phi$ is the sum of the left-most values of the central and satellite curves. 

The fixed HOD under-predicts the response of $N_g$ to $\sigma_8$, in particular for central galaxies. This is consistent with \cite{voivodic_response_2021}, who already noted that the HOD response to a large-scale fluctuation is not negligible. The fixed HOD does captures the response of the halo mass function, but it misses the change of the occupation at fixed halo mass, which is a large part of the $b_\phi$ response. 

\begin{figure}[ht!]
    \centering
    \includegraphics[width=\hsize]{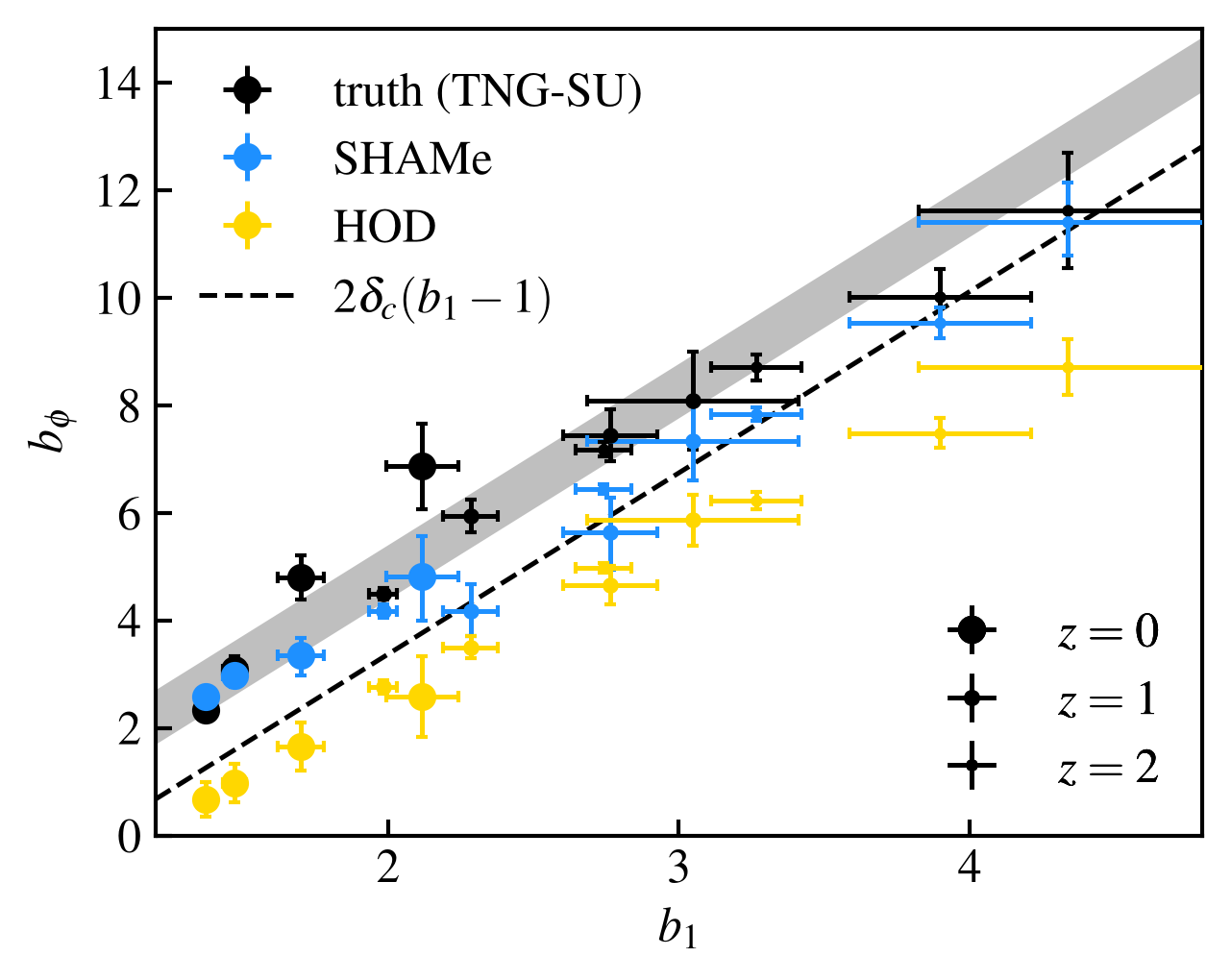}
    \caption{$b_{\phi}$ as a function of $b_1$, for the TNG-SU simulations (black), the SHAMe-SU prediction (blue) and the fixed-HOD prediction (yellow). Large dots, small dots, and crosses correspond to $z=0$, $1$, and $2$, respectively. The universality relation is displayed as a dashed line, and the grey band the relation with $p \in [0.4, 0.7]$ calibrated by  \citet{barreira_galaxy_2020} on the same simulations. For each sample, $b_1$ is measured by fitting a constant to $\xi_{gm}(s)/\xi_{mm}(s)$ measured in the TNG fiducial simulation in real space, in the range $s \in [15, 30] \, \Mpch$, using jackknife errors with $64$ jackknife regions. Error bars on $\bphi$ for the TNG-SU and HOD are jackknife errors, while SHAMe errors also include uncertainties on SHAMe parameters.}
    \label{fig:bphi-b1}
\end{figure}

Figure~\ref{fig:bphi-b1} shows $\bphi$ as a function of $b_1$, for the TNG-SU (black), SHAMe-SU (blue) and the fixed-HOD (yellow). We fit $b_1$ on the TNG-SU fiducial simulation, by measuring $\xi_{gm}(s)/\xi_{mm}(s)$ in real space in the range $s \in [15, 30] \, \Mpch$, using jackknife errors with $64$ jackknife regions. We display the universality relation $\bphi = 2 \delta_c (b_1 - 1)$ as a black dashed line, as well as the range $\bphi = 2 \delta_c (b_1 - p)$, $p \in [0.4,\,0.7]$ from \citet{barreira_galaxy_2020} as a grey band. We quantify the errors as the median over the 12 samples of $|b_\phi^{\rm pred}-b_\phi^{\rm TNG}|/b_\phi^{\rm TNG}$. The median is 10\% for SHAMe-SU, 38\% for the fixed HOD and 23\% for the universality relation.

The SHAMe-SU prediction is closer to the true value than both alternatives in all 12 samples. The fixed-HOD prediction is below the universality relation for every sample. At the highest $b_1$, SHAMe- approaches the universality relation, and the gain is not significant there.   

The SHAMe-SU prediction is compatible with the $p$-range proposed by \citet{barreira_galaxy_2020} for $b_1 < 2$, although they slightly under-predict the true $\bphi$ value for the highest $b_1$ points. We stress that the range $p \in [0.4,\,0.7]$ by \citet{barreira_galaxy_2020} was calibrated on these same TNG-SU simulations, so it is not an independent prediction for our samples. The SHAMe-SU predictions, in contrast, use only small-scale clustering and the SMF of the fiducial simulation. 

\subsection{Origin of the residual error}

We now explore where the SHAMe-SU residual error comes from. 

\begin{figure}[ht!]
    \centering
    \includegraphics[width=\hsize]{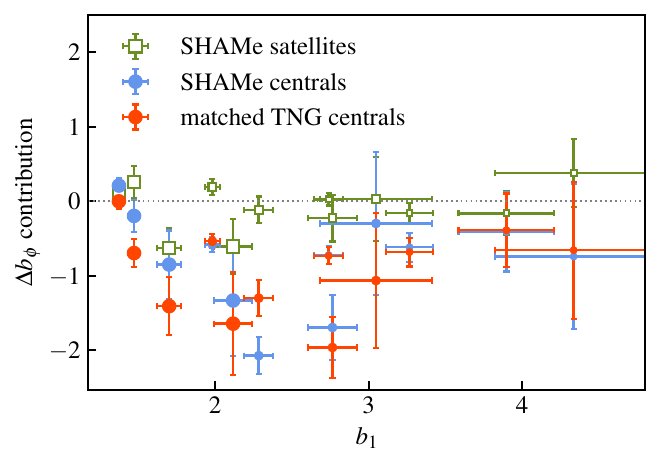}
    \caption{Contributions to the $\bphi$ error in our SHAMe prediction from central (blue points) and satellite (green squares) galaxies, as a function of $b_1$. Red points show the contribution from central galaxies when galaxy abundance is predicted from a RF trained on matched TNG subhalos. Large, medium and small markers correspond to the $z=0, 1, 2$ snapshots, respectively.}
    \label{fig:bphi-b1-errors}
\end{figure}

Figure~\ref{fig:hod_z0} indicates that it mostly originates from central galaxies near the mass threshold set by the $\mstell$ selection. We quantify this in figure~\ref{fig:bphi-b1-errors}, where we separate the contribution to the error in $\bphi$ arising from centrals and satellites. We compute central and satellite counts separately in both the TNG-SU and best-fit SHAMe-SU simulations, fit equation~\eqref{eq:bphi_line_fit} for each, and renormalize the resulting $\bphi$ by the total fiducial count. We define $\Delta \bphi = \bphi^{\rm SHAMe}-\bphi^{\rm TNG}$. 

To estimate the error on the difference between the SHAMe-SU mocks and the truth (TNG-SU), we repeat the fit for the same 125 leave-one-out regions in both sets and compute the jackknife covariance of $\Delta \bphi$. As we only show the best-fit SHAMe-SU prediction, these errors do not account for SHAMe parameters error. We see that the contribution of satellites to the error in $\bphi$ is mostly compatible with 0. Hence centrals account for the main contribution to the bias in the SHAMe $\bphi$ prediction. 

The bias in SHAMe predictions for $\bphi$ suggests that at least one of the assumptions we have made in section~\ref{sec:method} fails. Either the subhalo properties used in SHAMe do not carry the full $\sig$ dependence of $\mstell$, or the small-scale clustering does not constrain the galaxy--subhalo relation sufficiently well. To isolate these assumptions, we take advantage of the shared initial conditions between our TNG-mimic-SU and TNG-SU simulations. We match central subhalos in the three TNG-mimic-SU simulations to their TNG-SU counterparts\footnote{We cannot train directly on the TNG-SU subhalo properties because mass-accretion histories are not available for these simulations.} by matching their host halos, minimizing $\Delta(\vx)/(\Mpch) + \Delta(\log M_h)$ where $\Delta(\vx)$ is the Euclidean distance between the two halos and $\Delta(\log M_h)$ is the absolute difference in log mass. 

For all redshifts and densities, we train a RF to predict $\prob{\left. \mstell > \mstellcut \right| \xsub}$, using $\vpeak$ and the per-object bias as input features\footnote{The other subhalo properties used in SHAMe are only relevant for satellite galaxies.}. We then apply the trained RF to the three TNG-SU simulations. We train the RF jointly on all three simulations, without providing information about which separate universe each object belongs to, in order to reduce the training error. 

The resulting error in $\bphi$ is shown in red in figure~\ref{fig:bphi-b1-errors}. We find that even when learning $\prob{\left. \mstell > \mstellcut \right| \xsub}$ directly from the TNG simulations, using the same subhalo properties as in SHAMe, the resulting error in $\bphi$ is comparable to that of the SHAMe-SU prediction. Although both the matching procedure and the RF predictions have intrinsic errors, we have verified that including $\sig$ as an additional input feature allows us to recover the true $\bphi$ within jackknife uncertainties. This provides a consistency check that these errors are not responsible for the error in $\bphi$ shown in figure~\ref{fig:bphi-b1-errors}. 

This indicates that the small-scale clustering measurements are sufficiently informative to constrain the dependence of galaxy abundance on these subhalo properties, but that the properties themselves do not contain enough information to fully capture the response to $\sig$, particularly for central galaxies. A more accurate prediction of $\bphi$ may therefore require additional information about the mass-accretion history or environment of central subhalos.

\section{Conclusion}\label{sec:conclusion}

Data-informed priors on $\bphi$ would strengthen constraints on $\fnl$ from galaxy power spectrum analyses, without relying on relations between $\bphi$ and other bias parameters calibrated on simulations. In this work, we propose a route towards such priors. We fitted an extended SHAM model to small-scale clustering and propagated the posterior on its parameters to $b_\phi$. We validated this approach using the IllustrisTNG hydrodynamical separate-universe simulations. For 12 stellar-mass-selected galaxy samples at $z = 0$, 1 and 2, we compared the true $b_\phi$ with the value predicted by SHAMe.

Despite neither being developed for nor calibrated on PNG measurements, SHAMe predicts the true $b_\phi$ with a median error of 10\% using only the small-scale redshift space multipoles and the SMF of the fiducial TNG simulation. A fixed-HOD model, which neglects the response of the halo occupation to the long-wavelength potential, has an error of 38\%, and the universality relation of 23\%. Overall, SHAMe is closer to the true value than either of these two models in almost all 12 samples, since it captures the response to $\phi$ of the halo and subhalo populations relevant to a specific galaxy sample, including their correlation with the local environment --- these aspects are absent from the universality relation and the fixed HOD includes only the response of the halo mass function.

Our results also show that a good fit to small-scale clustering does not guarantee an accurate $b_\phi$: the inferred response also depends on the assumed galaxy--subhalo connection. For SHAMe, the residual error (10\% median, up to $\sim 30\%$) indicates that part of the response to $\phi$ is carried by subhalo properties that the model does not use. The fundamental origin of this limitation is not unique to SHAMe; it affects any galaxy-formation model.

Empirical models condition galaxy properties on a finite set of halo or subhalo properties, whereas hydrodynamical simulations determine them through their sub-grid prescriptions. Any part of the response to $\phi$ that these properties or prescriptions do not capture becomes a residual error in $b_\phi$. A fixed HOD is the extreme case, in which the occupation is assumed not to respond at all. Hydrodynamical simulations are not exempt either: variations of feedback prescriptions shift the $b_\phi(b_1)$ relation \citep{perez_impact_2026}, so a galaxy-formation model other than IllustrisTNG could predict a weaker or a stronger residual response. The discrepancy we find in TNG therefore measures the size of this term in one galaxy-formation model, not the error of our approach when applied to the real Universe. For real data, clustering alone cannot constrain the galaxy-formation-dependent part of $b_\phi$ beyond the response of halos and subhalos, so this part should be treated as a theoretical uncertainty.

Additional subhalo properties could reduce this galaxy-formation-dependent part. Examples are halo formation time, since at fixed $V_{\rm peak}$ older halos have had more time to form stars, and a more detailed description of the environmental dependence of stellar mass. We will explore these in an extended version of SHAMe. These new parameters are likely to be poorly constrained by small-scale clustering, but this is a positive feature rather than a drawback: they will broaden the $b_\phi$ posterior, so that the freedom of the model propagates into the prior. Since no empirical model or hydrodynamical simulation can guarantee the correct $b_\phi$ for the real Universe, the goal should be priors whose width reflects both the clustering constraints and our uncertainty about galaxy formation.

Several further steps are needed before this approach can be applied to data. Our analysis focuses on stellar-mass-selected galaxy samples, which provide a useful approximation to selections such as those of the Bright Galaxy Survey (BGS) or Luminous Red Galaxies (LRG) in DESI, for example. Applying this framework to samples selected according to other galaxy properties will require adapting the galaxy-formation model to account for the relevant selection. For instance, star-forming galaxies such as Emission Line Galaxies (ELGs) could be modelled using the SHAMe-SF framework developed by \citet{ortega-martinez_shame-sf_2024}, which explicitly accounts for star formation activity.

We have also fixed cosmological parameters. In a realistic analysis, cosmological and galaxy-formation parameters should be jointly constrained, and the resulting posterior on $\bphi$ would be marginalized over both. This is conceptually straightforward and could be achieved by combining separate-universe simulations spanning different cosmologies, but it would likely broaden the posterior on $\bphi$. In our analysis, the constraints on galaxy-formation parameters are limited by the finite volume of the separate-universe simulations $(205 \, \Mpch)^3$, which is considerably smaller than the volume probed by current surveys such as DESI. Using the much larger volume of observational data could therefore improve the constraints on galaxy-formation parameters and compensate for the additional uncertainty introduced by marginalizing over cosmology. Nevertheless, extending the analysis in this direction will require validating the accuracy of the pipeline at a higher level of precision.

Despite these limitations, our results demonstrate that empirical galaxy-formation models constrained by small-scale clustering can capture a substantial part of the response of galaxy formation to local PNG. This provides a promising avenue for obtaining data-informed priors on $\bphi$, and more generally other perturbative bias parameters, directly from galaxy clustering observations. Ultimately, such an approach would help exploit the information contained in both small and large-scale clustering to obtain more comprehensive cosmological constraints from current and future galaxy surveys.

\begin{acknowledgements}
We thank F. Schmidt for providing the TNG-SU simulations used in this work. We also thank E. Lizalde for his contribution to the project in its early stage.
MP and REA received support from grant PID2024-161003NB-I00 funded by MICIU/AEI/10.13039/501100011033. SC acknowledges support from the Ramón y Cajal fellowship (RYC2023-043783-I). SC also acknowledges support from grant PID2025-171262NB-I00, funded by MICIU/AEI/10.13039/501100011033 and by ERDF/EU. AMD is supported by the Agence Nationale de la Recherche (ANR) under grant No.~ANR-23-CPJ1-0160-01.
\end{acknowledgements}

\bibliography{bibtex/references}

\begin{appendix}
\nolinenumbers

\section{Box-size effects}\label{sec:boxsize}

\begin{figure*}
    \centering    
    \includegraphics[width=\hsize]{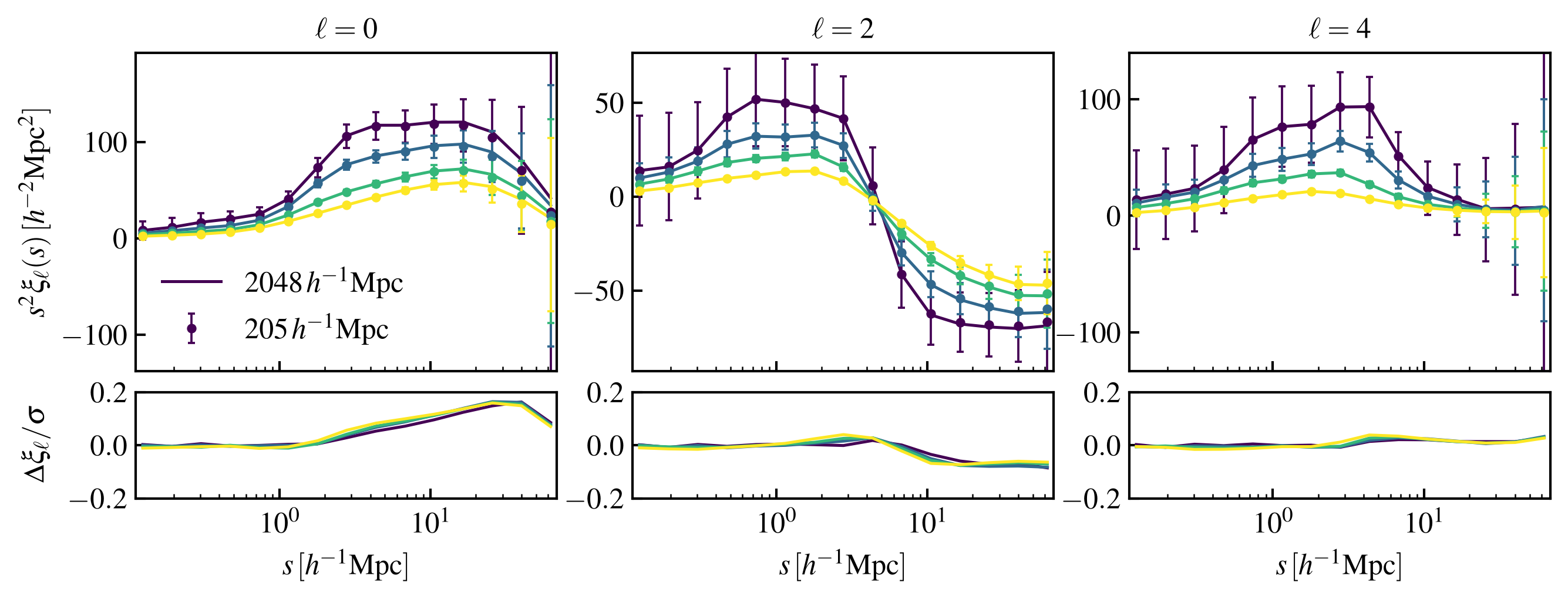}
    \caption{Effect of simulation box size on the correlation function multipoles. Lines show the multipoles computed on the $(2048 \, \Mpch)^3$ TheOne simulation at $z=2$, while points show the average of 1000 $(204.8 \, \Mpch)^3$ sub-boxes of the same simulation, with error bars equal to the scatter over the 1000 sub-boxes. Blue to yellow points show the four densities considered in this work, in the following order $\log (\nbar\, \nunits ) = -4, -3.5, -3, -2.5$. Bottom panel shows the difference between the sub-boxes average measurement and the large-box measurement, divided by the standard deviation over the 1000 sub-boxes.}
    \label{fig:boxsize}
\end{figure*}

We check the effect of missing large-scale modes in the TNG-SU simulation volume, $(205 \Mpch)^3$, by comparing the correlation function multipoles computed in \num{1000} sub-boxes from the $(2048 \, \Mpch)^3$ simulation used to compute the sample covariance (section~\ref{sec:data_covariance}). To circumvent non-periodic boundaries of the sub-volumes, we compute the cross-correlations between the sub-boxes and the full box. We normalize the correlation function estimator by the average density in the sub-box. Figure~\ref{fig:boxsize} shows the comparison between the correlation function multipoles measured in the full $(2048 \, \Mpch)^3$ box at $z=2$, and the average of the 1000 sub-box measurements. The sub-boxes are in agreement with the measurement within $20\%$ of the data error at the scales considered here. The effect is similar at the other redshifts. Therefore we ignore box-size effects in our analysis.

\section{SHAMe algorithm}\label{sec:SHAMe_algo}
In this appendix, we describe in more detailed the SHAMe algorithm we use to populate dark-matter-only simulations with stellar-mass-selected galaxies. The procedure is as follows:
\begin{enumerate}
    \item \label{orphan_item} orphan subhalos whose time since accretion --- i.e. time since they became a satellite of their current host halo --- exceeds their dynamical friction time scale $t_{\rm dyn}$ are considered to have merged with their central structure and are discarded from the simulation. Details on the computation of $t_{\rm dyn}$, which is normalized by $\logtmerger$ to effectively regulate the number of orphan galaxies, are given in \citet{contreras_flexible_2021b}. Additionally, the positions of the remaining orphan subhalos are shrunk towards the center of the host halo, depending on the time since accretion and $t_{\rm dyn}$;
    \item \label{sham_item} subhalos are ranked by decreasing $\vpeak$, and assigned a stellar mass to match the input SMF (here measured in the TNG fiducial simulation);
    \item for each subhalo, a new stellar mass is drawn from a Gaussian distribution around the previous value, with standard deviation $\sigmalum$, and step 3 is repeated with $\vpeak$ replaced by this new stellar mass value;
    \item then some level of assembly bias is added, following the $\abs{\fkp + \fkm}$ for centrals and $\abs{\fkp - \fkm}$ for satellites. Within narrow $\vpeak$ bins, stellar masses are reassigned to galaxies so that $\mstell$ is more (for high $\abs{\fkp \pm \fkm}$) or less (for low $\abs{\fkp \pm \fkm}$) correlated with the galaxy bias. This per-object bias is computed following the method from \citet{paranjape_halo_2018};
    \item finally, galaxies that have been satellites for longer than $t_{\rm rec}\betalum$ are assigned a zero $\mstell$, and are therefore effectively removed from any galaxy sample. $t_{\rm rec}$ is the lookback time to recombination, and $\betalum \in [0, 1]$. Hence $\betalum = 0$ means that all satellites have been destroyed, while $\betalum = 1$ means that they have all survived. Note that subhalos hosting a destroyed satellite galaxy still count as subhalos when computing $N_{\rm sub}$. Step \ref{sham_item} is then repeated again with $\mstell$ as the subhalo property to ensure that the resulting SMF still matches the observed one.
\end{enumerate}
We note that subhalos entering the sum in equation~\eqref{eq:ng_estimator} correspond to the number obtained after step~\ref{orphan_item}, and therefore depend on $\thetashame$ --- specifically $\logtmerger$.

\section{Covariance}\label{sec:covariance}
\subsection{Data covariance}\label{sec:data_covariance}

In this sub-section, we compare the covariance matrix obtained from jackknife resampling of the TNG-SU simulation volume (with 343 sub-regions), with an estimate from a larger dark-matter-only simulation with $2048 \, \Mpch$ side length and $N = 5600^3$ dark matter particles. We populate the latter with SHAMe galaxies, using the best-fit parameters obtained from fitting the clustering of the TNG-SU fiducial simulation. This fit was performed using a jackknife covariance, with SHAMe galaxies populated directly in the TNG-SU fiducial simulation at each model evaluation (i.e. without using an emulator). 

We then divide the large simulation volume into 1000 sub-boxes and compute the correlation function multipoles in each, using the mean number density measured within that sub-box to normalize the estimator. Since these sub-boxes do not satisfy periodic boundary conditions, we compute the cross-correlation between each sub-box and the full box to mitigate boundary effects. Given that the scales considered in this work are much smaller than the size of the sub-boxes, we expect the impact of this approximation on the covariance estimates to be negligible.

Although this estimate is derived from a single realization, its larger volume provides a more representative estimate of the true covariance of the target volume than the jackknife estimate. Hence, we adopt this estimate as our reference covariance. Additionally, to estimate the impact of varying cosmology and SHAMe model on the covariance, we compute the jackknife correlation function multipoles for a smaller version of the large box, which has a  $256 \, \Mpch$ side length and a comparable resolution ($N = 768^3$ particles). We then rescale the jackknife covariance by $(256/205)^3$ to match the expected covariance of the TNG volume. 

\begin{figure*}
    \centering    
    \includegraphics[width=\hsize]{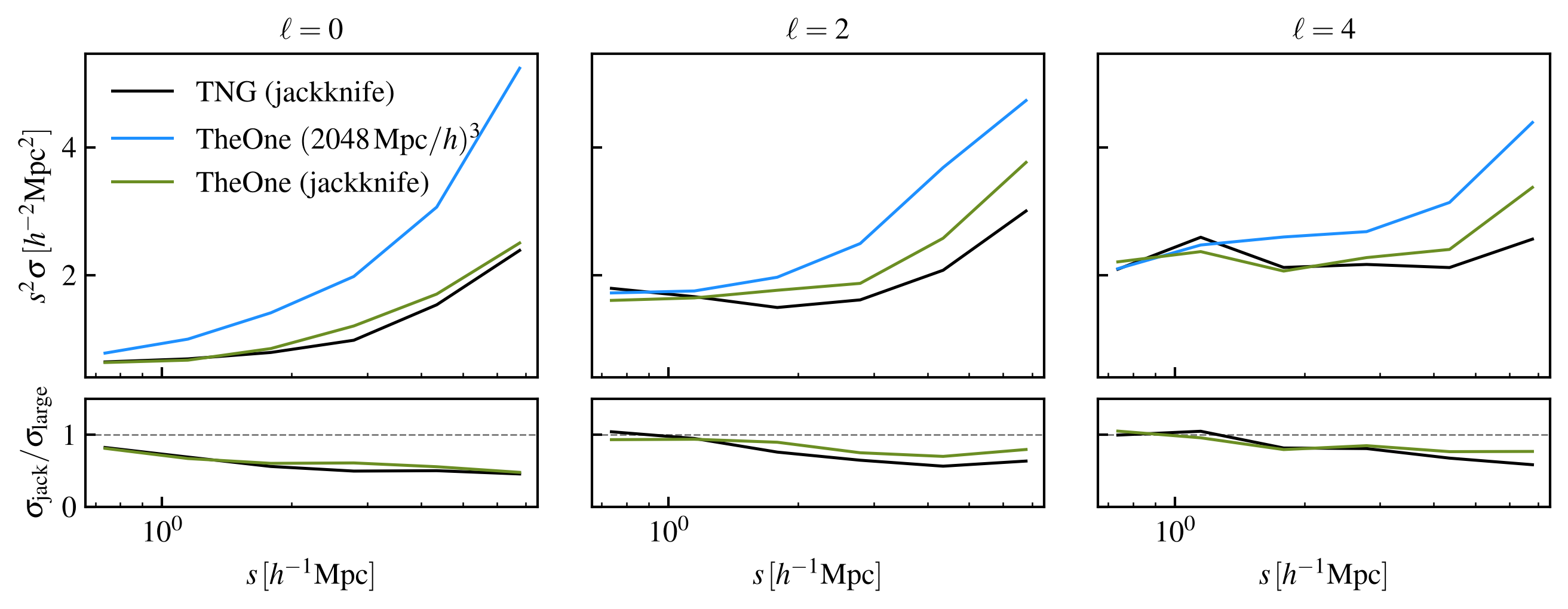}
    \caption{Comparison between the square root of the diagonal elements of three covariance estimates for the $z=0$, $\bar{n}_g  = 10^{-3} \, h^3 \mathrm{Mpc}^{-3}$ correlation function multipoles: a jackknife estimate (343 jackknife regions) from the TNG fiducial simulation (black), a similar-size $(256 \, \Mpch)^3$ dark-matter-only simulation populated with SHAMe galaxies (green), and a larger version of the same simulation ($(2048 \, \Mpch)^3$, blue). }
    \label{fig:sample_cov_diag}
\end{figure*}

\begin{figure*}
    \centering    
    \includegraphics[width=0.495\hsize]{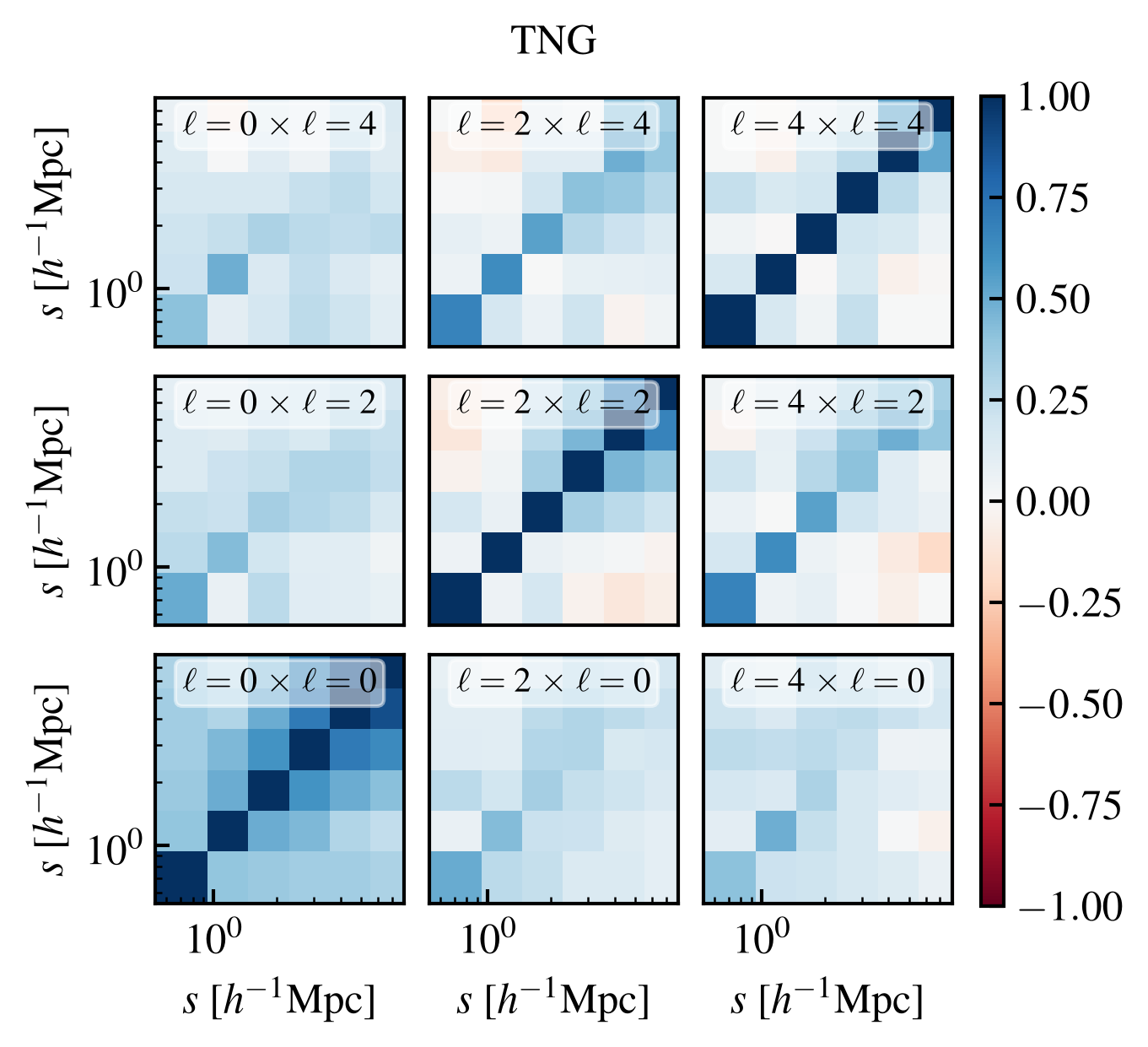}
    \includegraphics[width=0.495\hsize]{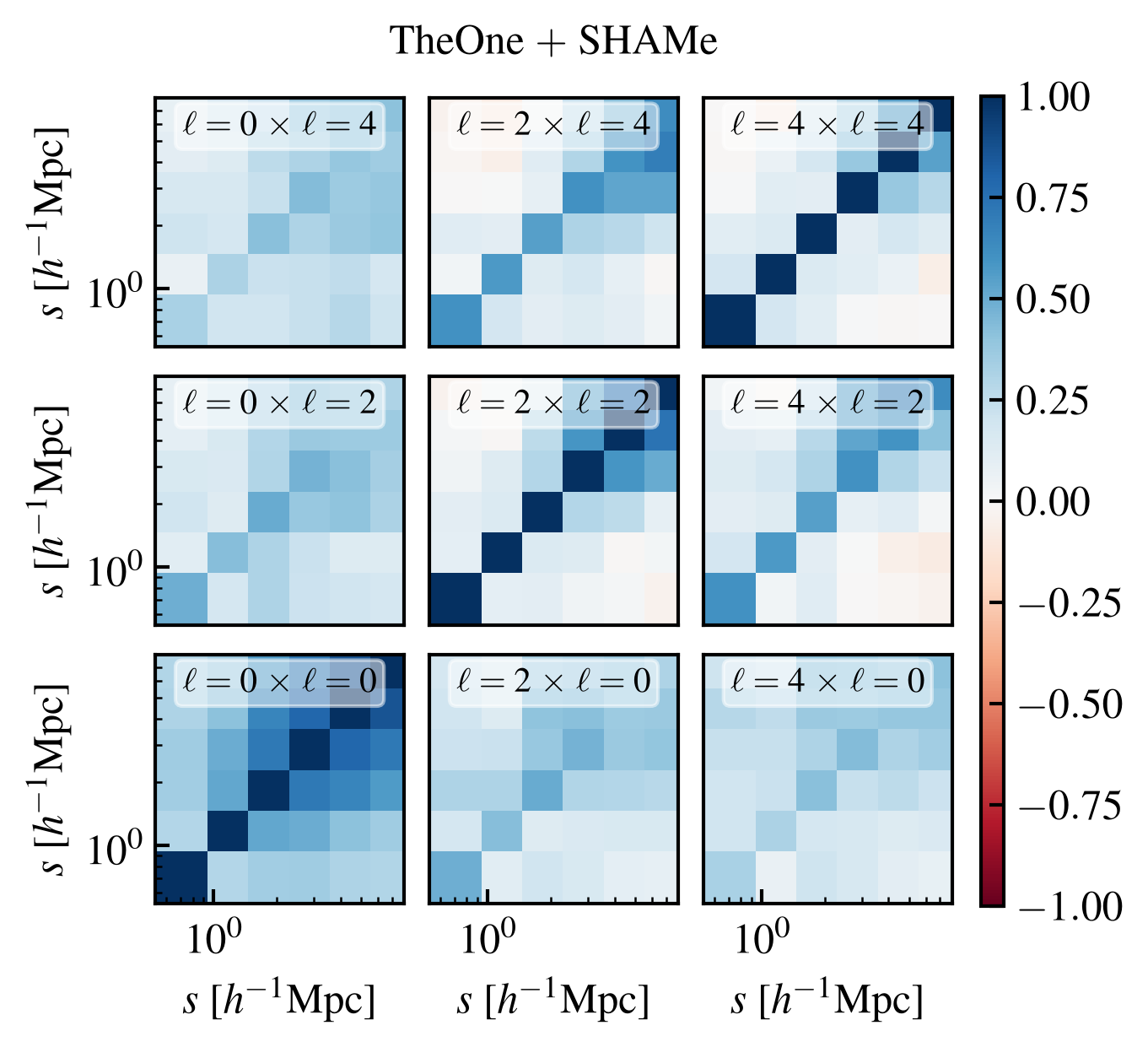}
    \caption{Jackknife correlation matrices from the TNG fiducial simulation (left) and TheOne dark-matter-only simulation populated with SHAMe galaxies (right). Both correspond to 343 jackknife regions. The galaxy population shown is is the same as in figure~\ref{fig:sample_cov_diag}, namely the $z=0$, $\bar{n}_g  = 10^{-3} \, h^3 \mathrm{Mpc}^{-3}$ sample.}
    \label{fig:sample_corrmatrix}
\end{figure*}

In figure~\ref{fig:sample_cov_diag}, we compare the square root of the diagonal elements of these three covariance estimates for the $z=0$, $\bar{n}_g  = 10^{-3} \, h^3 \mathrm{Mpc}^{-3}$ sample. We see that the jackknife estimate (green) is systematically smaller than the large-box estimate (blue) at almost all scales. This suggests that the jackknife covariance substantially underestimates the true covariance. However, the comparison between the jackknife estimates for both TNG and TheOne + SHAMe simulations agree reasonably well, indicating that the impact of the change in cosmology on the errors is small. While we do not show them, this observation remains consistent across all redshifts and densities considered in this work. Moreover, as we see in figure~\ref{fig:sample_corrmatrix}, the correlation matrices of the TNG and (small) TheOne + SHAMe simulations show similar structures. Therefore, we use the covariance from the $2048 \, \Mpch$ TheOne + SHAMe simulation to estimate the sample errors on the TNG-SU measurements. We use the same covariance for all three amplitudes.

\subsection{SHAMe error}\label{sec:shame_error}

\begin{figure*}
    \centering    
    \includegraphics[width=\hsize]{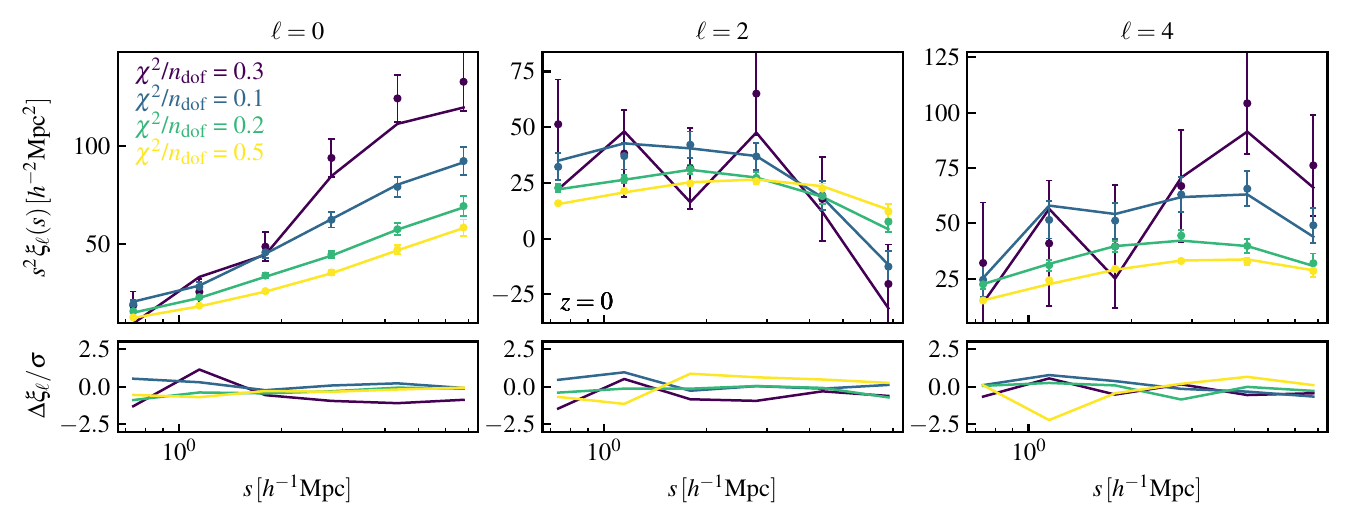}
    \caption{Measurements of the $\xi$ multipoles from the TNG-SU simulations at $z=0$ (points), compared to the prediction from the best-fit SHAMe parameters applied to the TNG-mimic-SU simulations (lines). Blue to yellow lines show the four densities considered in this work, in the following order $\log (\nbar\, \nunits ) = -4, -3.5, -3, -2.5$. The best-fit SHAMe parameters were obtained by generating SHAMe mocks on-the-fly and fitting them with a PSO optimizer. Bottom panel shows the residuals between the measurements and the model, normalized by the simulation error computed from the large-box estimator.}
    \label{fig:direct_bestfit_z0}
\end{figure*}

\begin{figure*}
    \centering    
    \includegraphics[width=\hsize]{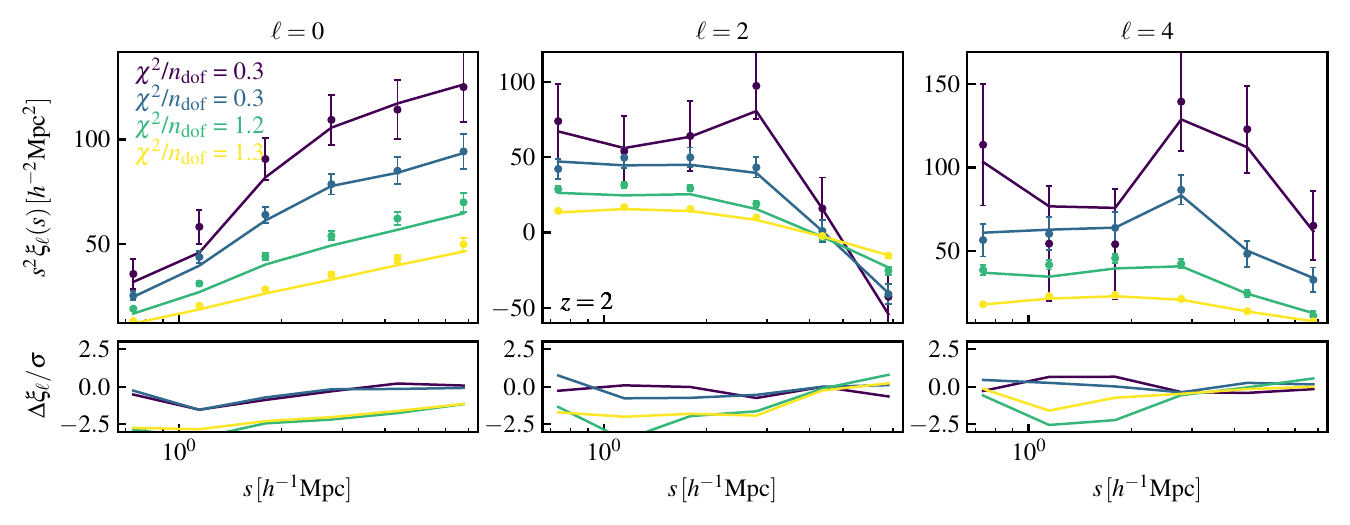}
    \caption{Same as figure~\ref{fig:direct_bestfit_z0}, for $z=2$.}
    \label{fig:direct_bestfit_z2}
\end{figure*}

To isolate the error of the SHAMe model from the emulator errors, we populate the TNG-mimic-SU fiducial simulation with SHAMe galaxies and measure their small-scale correlation function on-the-fly. We fit SHAMe parameters with a PSO optimizer (section~\ref{sec:covariance}). To reduce noise in the measurements, we average the correlation functions over four SHAMe seeds at each optimization step. We estimate the covariance matrix using jackknife resampling. Figures \ref{fig:direct_bestfit_z0} and \ref{fig:direct_bestfit_z2} show the best-fit SHAMe predictions, averaged over four seeds, compared to the fiducial TNG-SU simulation at $z=0$ and $z=2$, respectively. The bottom panels show the residuals normalized by the large-box covariance estimate (see section~\ref{sec:covariance}). The fits yield $\chi^2/n_{\rm dof}$ values well below 1 for most samples, with the exception of the two highest-density samples at $z=2$, for which the values are slightly above 1. At $z=1$ (not shown), all samples have $\chi^2/n_{\rm dof}<0.6$. Overall, this test indicates that SHAMe provides a good description of the small-scale clustering of the TNG-SU samples, although some residual discrepancies remain. We do not include an additional SHAMe modelling-error contribution in our covariance, as we lack a reliable way to quantify it.

\subsection{Emulator covariance}\label{sec:emu_cov}

\begin{figure*}
    \centering    
    \includegraphics[width=\hsize]{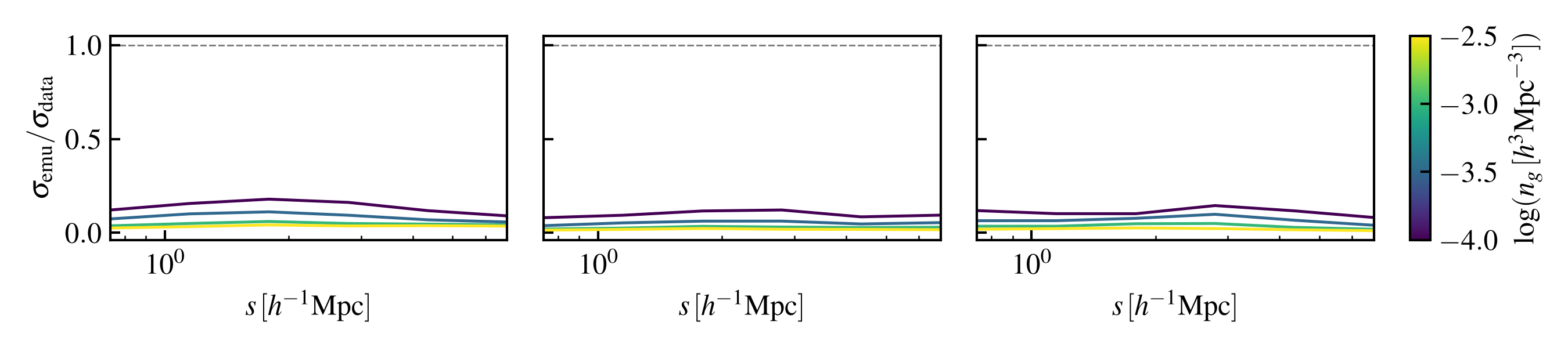}
    \caption{Square root of the diagonal elements of the emulator covariance, for all $\nbar$ values, over $z=0$ data errors from the large-box estimator.}
    \label{fig:emu_error}
\end{figure*}

For each $\nbar$ value, the emulator covariance is estimated by taking the covariance of the subset of the 800 test simulations within the bin $[\log \nbar - 0.3, \log \nbar + 0.3]$, where $\nbar$ is in $\nunits$. This corresponds to 121, 222, 221 and 222 samples for our four densities $\log(\nbar [\volunits]) = -4, -3.5, -3, -2.5$, respectively.
As the emulator error exhibits no strong redshift dependence, we do not apply any cut in redshift to these test sets. We estimate the emulator error as the covariance of the differences between the emulator prediction and the measured $\xi_{\ell}$ over the sample. The square root of the diagonal elements of this covariance, compared to the data error at $z=0$, are shown in figure~\ref{fig:emu_error}. The emulator error is below 20\% of the data error for all samples. We include it in our total covariance for completeness.

\subsection{Scaling covariance}\label{sec:scaling_cov}

\begin{figure*}
    \centering    
    \includegraphics[width=\hsize]{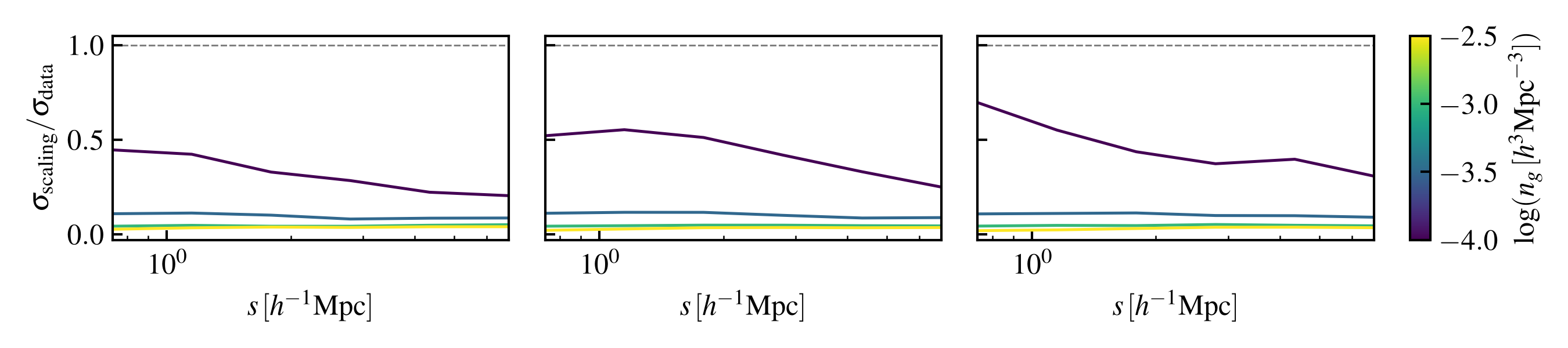}
    \caption{Square root of the diagonal elements of the scaling covariance at $z=0$, for all $\nbar$ values, over $z=0$ data errors from the large-box estimator.}
    \label{fig:scaling_error}
\end{figure*}

Similarly to \cite{contreras_milleniumtng_2023}, we estimate the error of the scaling technique by computing the residuals between rescaled and target simulations from a suite of 33 paired simulations with cosmologies distributed following a latin hypercube. For each pair, we apply the SHAMe algorithm with 10 sets of parameters randomly drawn from the uniform priors presented in table~\ref{tab:emu_cosmo_params}. For each set of parameters, we compute the average over 10 SHAMe seeds to eliminate shot noise and isolate the scaling error. The scaling error for all densities at $z=0$ is shown in figure~\ref{fig:scaling_error}, and is below $\sim 60\%$ of the data error for all densities. We have check that for the other redshifts (not shown here) the scaling error is similar or smaller than at $z=0$. We also include it in our total covariance.

\section{Goodness of fit for all densities}\label{sec:residuals_all}

\begin{figure*}
\centering    
\includegraphics[width=\hsize]{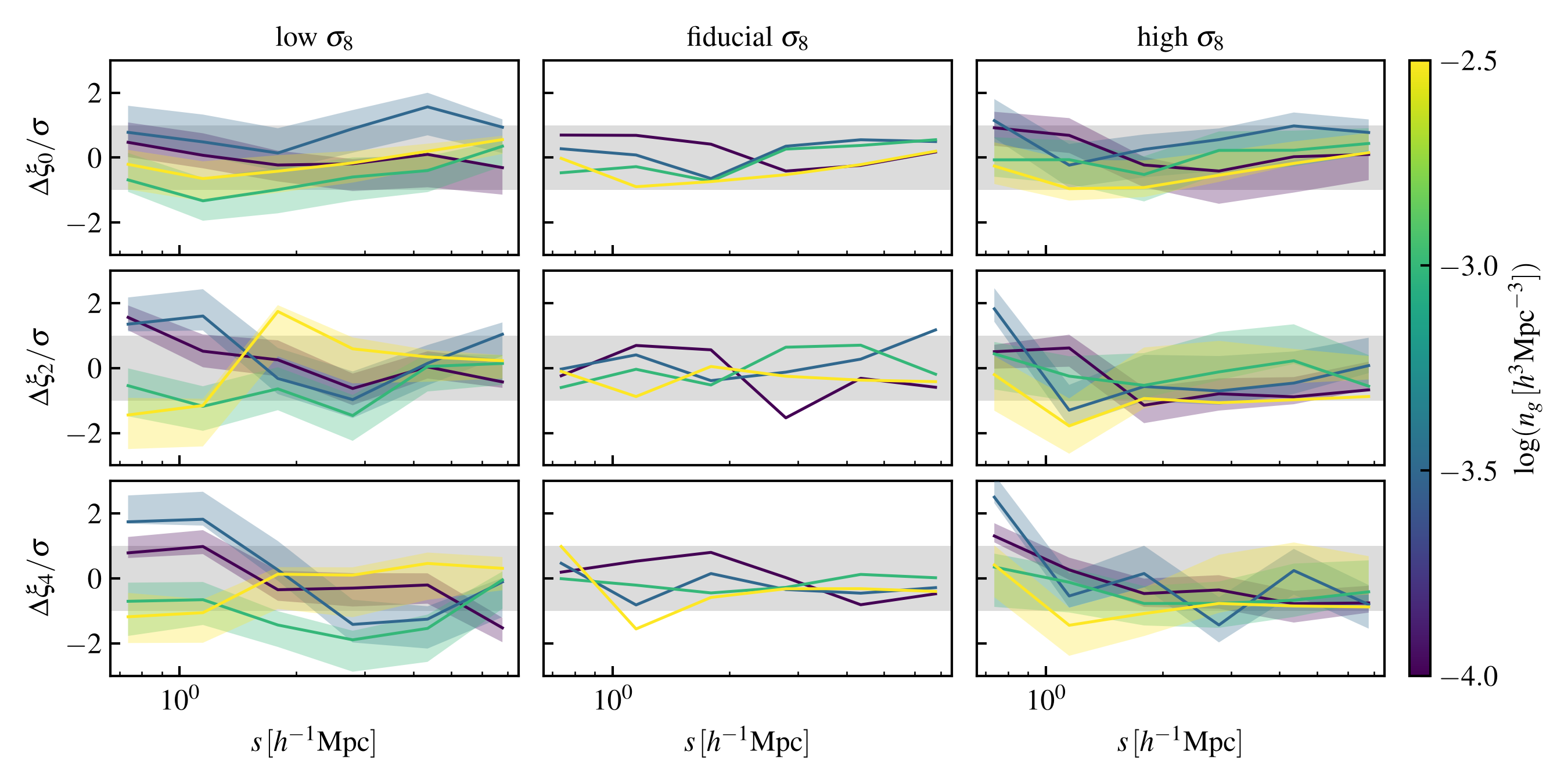}
  \caption{Residuals of the correlation function multipoles $\left( \xi_{\ell, \rm emu}(s) - \xi_{\ell,\rm data}(s) \right) / \sigma$ for all considered densities at $z = 0$. Solid lines show the residuals between the TNG-SU simulations and the fiducial best-fit model. For the low and high amplitudes, shaded areas show the per-point 68\% credible intervals computed from the sampled SHAMe parameters.}
\label{fig:residuals_emu_z0}
\end{figure*}

\begin{figure*}
\centering    
\includegraphics[width=\hsize]{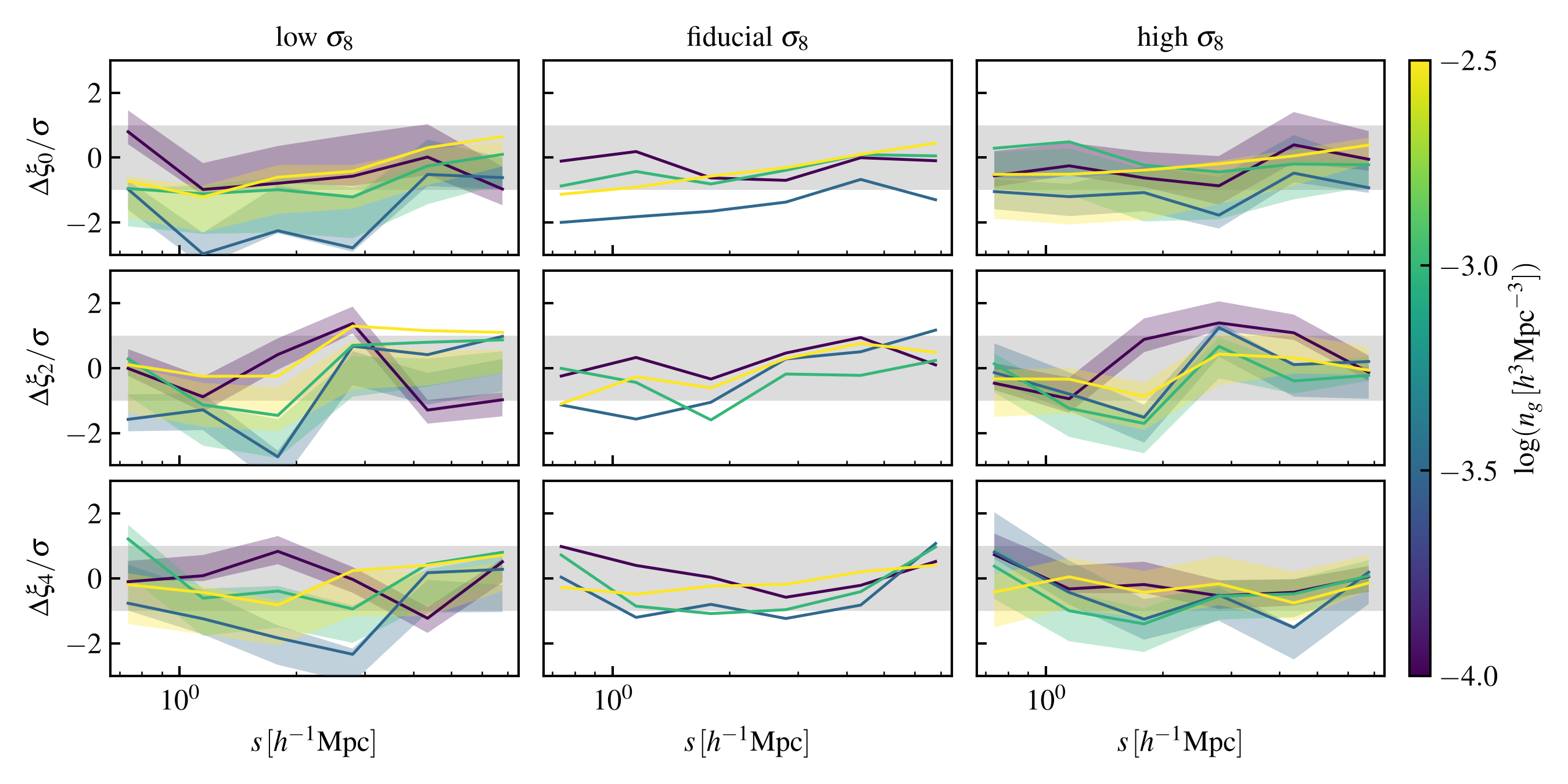}
  \caption{Same as figure~\ref{fig:residuals_emu_z0}, at $z=1$.}
\label{fig:residuals_emu_z1}
\end{figure*}

\begin{figure*}
\centering    
\includegraphics[width=\hsize]{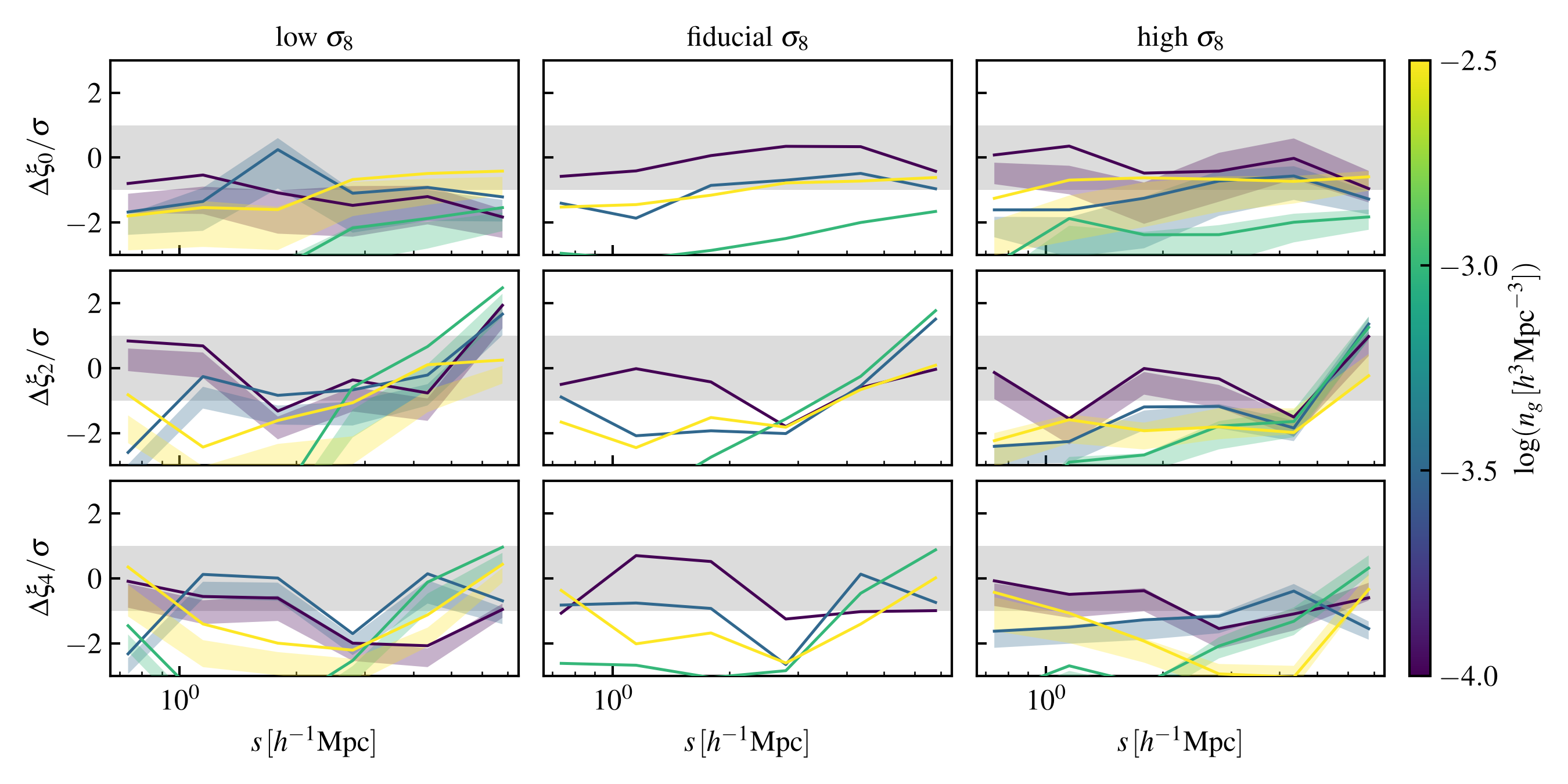}
  \caption{Same as figure~\ref{fig:residuals_emu_z0}, at $z=2$.}
\label{fig:residuals_emu_z2}
\end{figure*}

In figures \ref{fig:residuals_emu_z0} \ref{fig:residuals_emu_z1}, and \ref{fig:residuals_emu_z2}, we show the residuals between the emulator prediction evaluated at the fiducial best-fit parameters and the measurements from the TNG-SU for all densities, at redshifts $z=0$, $z=1$ and $z=2$, respectively. For the low and high amplitudes, we additionally display the $68\%$ credible intervals of the fiducial best-fit model, showing that the clustering of the separate universes is compatible with the fiducial best-fit prediction. Although the emulator prediction shows stronger deviation to the truth at $z=2$, all fiducial fits have a $\chi^2/n_{\rm dof}$ value below 1.

\section{Robustness to a misspecified SMF}\label{sec:misspecified_smf}

Since galaxies are selected in the fiducial simulation to match a target number density $\nbar$, and SHAMe relies on ranking subhalos, the clustering of the resulting SHAMe mock is largely insensitive to the specific shape of the input SMF. However, our procedure for predicting $\bphi$ relies on the relationship between the subhalo properties and whether a subhalo hosts a galaxy above the stellar mass threshold. This relationship, and hence the predicted number counts in the separate universes and the resulting $\bphi$, may depend on the assumed SMF.

\begin{figure*}
\centering    
\includegraphics[width=\hsize]{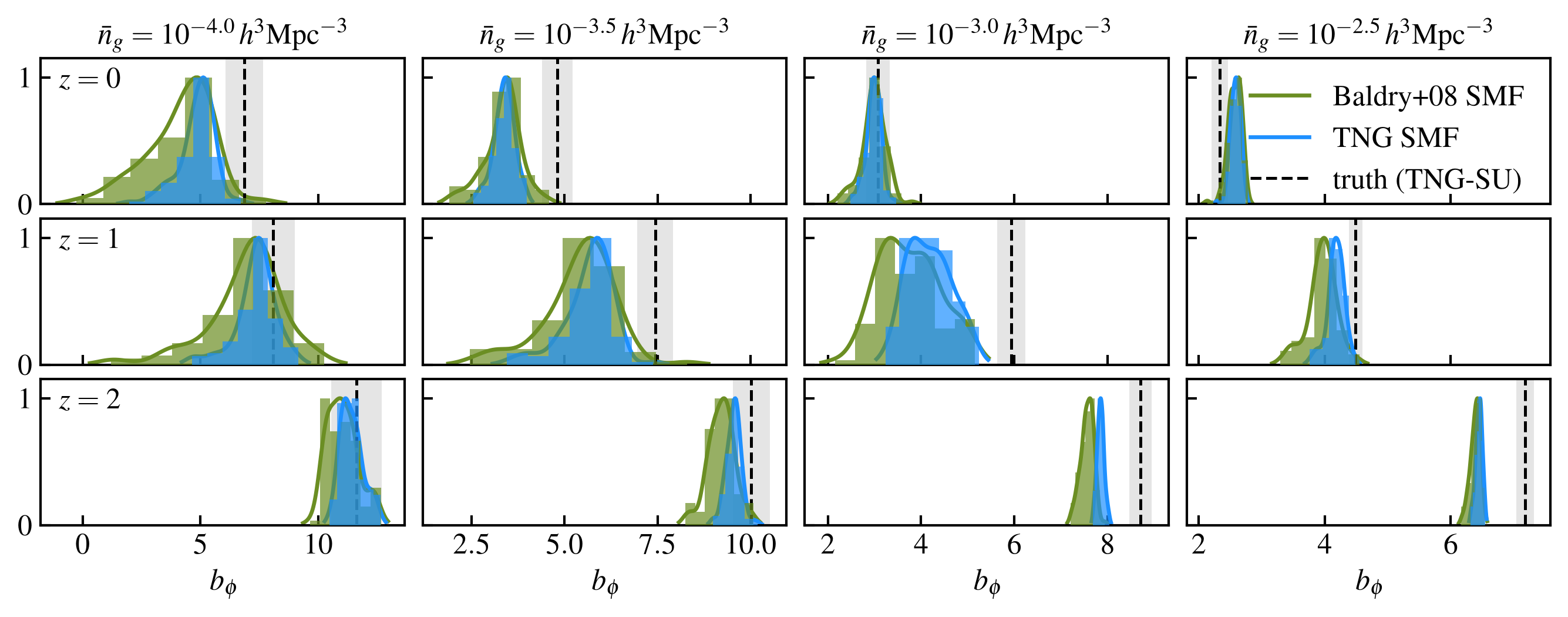}
  \caption{Posteriors on $\bphi$ obtained when using a misspeficied SMF from \cite{baldry_on_2008} in SHAMe (green), compared to the prediction obtained with the true TNG-SU fiducial SMF (blue). Vertical dashed lines correspond to the best-fit SHAMe parameters in both cases. The black vertical dashed line is the truth, with grey error showing the jackknife errors.}
\label{fig:misspecified_smf}
\end{figure*}

In this section, we test the robustness of our $\bphi$ predictions to SMF misspecification. To this end, we repeat the pipeline described in section~\ref{sec:method}, replacing the measured fiducial TNG-SU SMF with the theoretical prediction from \citet{baldry_on_2008}. The resulting $\bphi$ posteriors are shown in green in figure~\ref{fig:misspecified_smf}. The vertical dashed lines indicate the $\bphi$ predictions obtained using the best-fit SHAMe parameters in the original analysis (with the TNG-SU SMF) and in the test case (with the SMF from \citet{baldry_on_2008}). While some deviations are observed for the samples with the largest discrepancies between the predictions and the reference values, the results remain consistent between the two SMF choices for samples with accurate $\bphi$ predictions. This suggests that our $\bphi$ predictions are reasonably robust to SMF misspecification in the regime where the SHAMe model provides accurate predictions.

\end{appendix}

\end{document}